\documentclass[3p,times,12pt]{elsarticle}
\journal{J Theoretical Biology}

\usepackage{color}
\usepackage{xcolor}
\usepackage{soul}
\usepackage{mathcomp} 
\usepackage{amsmath}
\usepackage{amsfonts}

\usepackage{graphicx}
\let\hat=\widehat
\let\tilde=\widetilde
\newcommand{\ds}{\displaystyle}
\newcommand{\beq}{\begin{equation}}
\newcommand{\eeq}{\end{equation}}
\newcommand{\beqa}{\begin{eqnarray}}
\newcommand{\eeqa}{\end{eqnarray}}
\newcommand{\nn}{\nonumber}
 
\newcommand{\ep}{\epsilon} 
\newcommand{\ee}{\mbox{e}}

\newcommand{\ro}{\varrho}

\newcommand{\pad}[2]{\frac{\partial #1}{\partial #2}}
\newcommand{\padd}[2]{\frac{\partial^2 #1}{\partial {#2}^2}}
\newcommand{\paddd}[3]{\frac{\partial^2 #1}{\partial {#2} \partial {#3}}}

\newcommand{\mfrac}[2]{\mbox{$\frac{#1}{#2}$}}
\newcommand{\rec}[1]{\mfrac{1}{#1}}
\newcommand{\half}{\rec{2}}
\newcommand{\dd}{{\rm d}}

\newcommand{\sech}{\mbox{sech}}
\newcommand{\erfc}{\mbox{erfc}}

\newcommand{\ol}[1]{\overline{#1}}
\newcommand{\btheta}{\boldsymbol\theta}
\newcommand{\balpha}{\boldsymbol{\alpha}}
\newcommand{\bbeta}{\boldsymbol{\beta}}
\newcommand{\bdelta}{\boldsymbol{\delta}}
\newcommand{\bw}{\mathbf{w}}

\renewcommand{\theequation}{\arabic{section}.\arabic{equation}}
\newcommand{\qb}[1]{{#1}}

\normalsize

\begin{document}
%----------------------------------------------------------------------------

\begin{frontmatter} 

\title{{\bf  Analysis of Genotype-Phenotype Association using 
Genomic Informational Field Theory (GIFT)}} 

\author{Jonathan AD Wattis$^1$,  
Sian M Bray$^2$, 
Panagiota Kyratzi$^{1,3}$  
Cyril Rauch$^3$,  
  \\ 
$^1$ Centre for Mathematical Medicine and Biology, School of Mathematical Sciences,  \\ 
University of Nottingham, University Park, Nottingham NG7 2RD, UK \\ 
Jonathan.Wattis@nottingham.ac.uk \\ 
$^2$ School of Life Sciences, University Park, Nottingham, NG7 2RD, UK \\ 
Sian.Bray@nottingham.ac.uk \\
$^3$ Vetinary Academic Building, Sutton Bonington Campus, \\ 
University of Nottingham, Sutton Bonington, Leicestershire, LE12 5RD, UK \\
Cyril.Rauch@nottingham.ac.uk\\[-4ex]
}

\date{{ 11 February, 2022, revised, 25 May, 2022}} 

\normalsize

%\maketitle
%----------------------------------------------------------------------------

\begin{abstract}
We show how field- and information theory can be used to quantify 
the relationship between genotype and phenotype in cases where 
phenotype is a continuous variable.  
Given a sample population of phenotype measurements, from 
various known genotypes, we show how the ordering of phenotype data 
can lead to quantification of the effect of genotype.  
This method does not assume that the data has a Gaussian distribution, 
it is particularly effective at extracting weak and unusual dependencies 
of genotype on phenotype.  
However, in cases where data has a special form, (eg Gaussian), 
we observe that the effective phenotype field has a special form. 
We use asymptotic analysis to solve both the forward and reverse 
formulations of the problem. 
We show how $p$-values can be calculated so that the significance 
of correlation between phenotype and genotype can be quantified. 
This provides a significant generalisation of the traditional methods 
used in genome-wide association studies GWAS. 
We derive a field-strength which can be used to deduce how the 
correlations between genotype and phenotype, and their impact 
on the distribution of phenotypes. 
\\[2ex]
\\[2ex]
\noindent{\bf Highlights:} 
\begin{itemize}
\item 
new method for quantifying relationship between genotype and continuous phenotype
\item 
statistical significance can be calculated via explicit expressions for $p$-values
\item  
method makes no assumption on shape of distribution data 
\item 
forward and inverse problems solved explicitly for the case of weak gene effect 
\end{itemize}
\end{abstract}

\end{frontmatter} 

%{{
%\noindent{}JTB Editor ?  F. Guichard, Montreal, Quebec, Canada 
%Ruth Baker, Oxford,  % ruth.baker@maths.ox.ac.uk  

%\\ Referee Suggestions: 
%Anne Skeldon, Surrey % a.skeldon@surrey.ac.uk 
%Ed Codling, Essex, ecodling@essex.ac.uk
%Radek Erban, Oxford, % erban@maths.ox.ac.uk 
%Tony Shardlow, bath % T.Shardlow@bath.ac.uk 
% Byron Morgan, Kent, % B.J.T.Morgan@kent.ac.uk

%Matt Keeling, Warwick, % M.J.Keeling@warwick.ac.uk 
%Vassili Kolokoltsov, Warwick % v.kolokoltsov@warwick.ac.uk 
%Hermanus Draisma, Surrey, %h.draisma@surrey.ac.uk 
%Jonathan Bartlett, % J.W.Bartlett@bath.ac.uk }} 

\normalsize

%----------------------------------------------------------------------------
\section{{Introduction} \label{sec-intro}}
\setcounter{equation}{0}

Explaining the causes of phenotypic variation has been an 
aim of natural science since its inception.  In more recent times, 
determining the extent to which genetic variation, as opposed to 
environmental factors, cause variation has been a heated topic 
of discussion.  With the massively increased availability of data 
in last few years, it is now possible to use statistical tools to 
quantify the effect of individual genes on phenotype.  

The most commonly used tool in this field is Genome-Wide 
Association Studies (GWAS) \cite{manolio,pearson}.  This method 
typically identifies correlations between a genetic variant and 
the presence of a particular condition or disease. 
These methods make use of only basic features of the distribution 
of phenotypes for each genotype, such as mean and variances 
of subpopulations, and the differences in means. 
The term `genetic variant' means Single Nucleotide Polymorphisms 
(SNPs) - which is the alteration of a single nucleotide (A, C,G, or T) 
in the DNA.  GWAS is then used to produce `Manhattan plots' that  
show $p$-values which show the association between the point mutation 
and likelihood of having a particular disease. This corresponds to a discrete 
phenotype, as individuals either {\em have} or {\em do not have} 
any particular disease. This approach has led to the identification 
of groups of genes involved in various diseases \cite{manolio,pearson}

In this paper, we address the more complicated scenario of a continuous 
distribution of phenotype values, for example, height, weight, BMI. 
The aim of this paper is to provide a theoretical framework to 
understand the relationship between phenotype and genotype.  
To establish a basic theory, we consider a single phenotype, 
for example, height or weight and a single gene.  
We then assume that the genetic state of each individual is known, 
the information contained in the sequence of genetic states is analysed. 
In Section \ref{deriv-sec} we derive an algorithm to calculate 
statistics from the observations of phenotype and genotype. 
The algorithm considers a sample taken from the population, and 
ranks individuals from the sample according to their phenotype 
(eg putting them in height order), to form an ordered list.  
We then calculate statistical values to quantify the 
significance of various outputs in Section \ref{pval-sec}. 
The underlying mathematical basis of the algorithm is 
derived in Section \ref{math-sec}, where we explain 
a generalisation of Shannon's Information theory \cite{Shannon}. 
We postulate an effective genotype field whose effect is to account for 
any skewness in the phenotype distribution;  we then make use of 
variational calculus to compute this field from the observed 
genotype-phenotype data sequence. \qb{We describe the resulting model 
as {\em GIFT}, that is, Genomic Informational Field Theory}. 

The model gives rise to two problems: the forward and inverse: 
the forward problem corresponds to the determination of location 
probabilities from theorised field strengths, whilst the inverse problem 
refers to obtaining field strengths from observed location probabilities. 
Since the inverse problem is simpler to solve, we consider that first, 
in Section \ref{inv-sec}. The forward problem is addressed in section   
\ref{weak-sec} using asymptotic analysis to solve the case of a 
weak interaction between genotype and phenotype. 
In Section \ref{pval-sec}, we perform more detailed calculations 
on the range of possible arrangements of genotypes in the list so that we 
can assign $p$-values to any particular observed outcome. 
Results of numerical simulations which illustrate how the method 
works are presented in Section \ref{num-sec}. 
Finally, conclusions are drawn and discussed in Section \ref{conc-sec}, 
whilst the appendices contains some of the lengthier mathematical derivations, 
in particular, the variational derivatives, \ref{AppA-ELderiv-sec},  
a master-equation approach \ref{res-sec} which complements the calculation 
of $p$ values in Section \ref{pval-sec}.  A final appendix (\ref{pheno-dep-sec}) 
shows how the minor modifications required if one wanted to plot results against 
actual phenotype value rather than aganst a position in the ordered list.

%----------------------------------------------------------------------------
\section{{Statistical algorithm}  \label{deriv-sec}}
\setcounter{equation}{0}

%---------------------
\subsection{{\bf Experimental setup \& observable data}}

We assume that a sample of individuals has been taken, and for each 
individual, there is genetic data available and a phenotype measurement 
has been made.  
We use $N$ to denote the size of the population sample, and enumerate 
individuals using $j$ where $1 \leq j \leq N$. 
We denote the phenotype by $\Omega$, which we assume is a 
continuous variable, that is, $\Omega \in \mathbb{R}$, 
and we label this data by individual, $j$, thus $\Omega(j)$. 
We assume that the gene occurs in one of three states, 
as occurs in diploid organisms.   For example, the case of 
two dominant alleles (AA) will be denoted `$+1$', the 
heterozygous state (Aa) is denoted by `0' and the 
homozygous state of two recessive alleles (aa) by `$-1$'.
We use $N_+,N_0,N_-$ for the numbers in each genetic state,  
then we have $N = N_+ + N_0 + N_-$. 

The method is based on the comparison of two arrangements of individuals.  
First, we consider the {\em ordered} state in which the individuals are 
arranged in increasing phenotype measurements, that is 
\beq
\Omega(1) < \Omega(2) < \Omega(3) < \ldots < \Omega(N) . 
\label{Om-order}
\eeq
For example, if our sample are horses, and the phenotype is height, then 
we can envisage this as allocating horses to paddocks based on their height: 
the shortest horse to the first paddock $j=1$, the second shortest horse to 
paddock $j=2$, etc, and paddock $j=N$ to the tallest horse.    
This allocation is based purely on phenotype and there is no explicit 
influence of genotype on the arrangement. 

Now we assume that the genetic state of each individual is known, 
that is, for each subject $1 \leq j \leq N$, we know whether it is $+1,0,-1$.  
We denote this state by $\gamma_j$ where, for each $j$, $\gamma_j$ 
takes one of the values $q \in\{+1,0,-1\}$.   We thus construct a sequence 
$\Gamma$ of genetic states given by 
\beq
\Gamma = ( \gamma_1, \gamma_2, \gamma_3, \ldots \gamma_N ) , 
\label{Gamma-def}
\eeq
where the order is important, since $\gamma_j$ corresponds to 
phenotype $\Omega(j)$.  
As an example, $\Gamma=(+1,+1,0,-1,0,+1,0,0,-1,-1,-1)$ represents a 
sample of $N=11$ individuals, $N_+=3$ of which have the +1 
genetic state (AA), $N_0=4$ are of heterozygous (`0'=Aa) and 
$N_- = 4$ are recessive and homozymous (`$-1$'=`aa').  
This list of information, $\Gamma$, (\ref{Gamma-def}) is the key quantity 
which we wish to analyse to determine the strength of genetic on phenotype. 

Clearly if the first $N_+$ of these states are all $\gamma_j=+1$, 
and the next $N_0$ are all $\gamma_j=0$, and the remaining $N_-$ 
are all $\gamma_j=-1$, then the genotype has a strong influence 
on the phenotype.  However, if the sequence $\Gamma$ 
appears random, then the genotype and phenotypes have no correlation 
and we can confidently claim that the gene has no influence on phenotype.  
Between these two extremes, there are the real-life cases where there is 
some correlation, between genotype and phenotype, without the magnitude 
of the effect being clear.   We propose to use information theory to find 
the strength and form of the relationship. 

The second allocation method we refer to as a completely {\em random} 
configuration or, rather, the average over all possible arrangements of 
individuals to positions in the list.   In our example, horses are allocated to 
paddocks with no influence of phenotype or genotype, so there is a probability 
of a paddock being occupied by a horse of a particular genetic state, 
and this probability is the same for all paddocks. 
 
We then compare the {\em actual} ordering of genotypes by phenotype 
(\ref{Om-order})--(\ref{Gamma-def}) with the random configuration. 
From $\Gamma$, we construct the cumulative distribution of 
homozygous or heterozygous states as follows. 
We define $W_+(j)$ to be the number of `$+$'-states occurring 
in the first $j$ individuals, that is, in the sub-list 
$(\gamma_1,\gamma_2,\ldots \gamma_j)$. Similarly, $W_0(j)$ 
is the number of 0-states in the first $j$ individuals, and $W_-(j)$ 
as the number of `$-$' states in the first $j$ individuals.  
Using the Kronecker $\delta$ symbol, defined by $\delta_{i,j}=1$ 
if $i=j$ and $\delta_{i,j}=0$ otherwise, the cumulative distributions 
can be expressed as 
\begin{align}
W_+(j) = & \sum_{i=1}^j \delta_{1,\gamma_i} = 
\mbox{number of $+1$'s in the first $j$ elements of the list $\Gamma$} , \nn\\ 
W_0(j) = & \sum_{i=1}^j \delta_{0,\gamma_i} = \mbox{} , 
\mbox{number of $0$'s in the first $j$ elements of the list $\Gamma$} , \nn\\ 
W_-(j) = & \sum_{i=1}^j \delta_{-1,\gamma_i} = \mbox{} . 
\mbox{number of $-1$'s in the first $j$ elements of the list $\Gamma$} , 
\label{WWq-def}
\end{align}
For completeness, we extend these definitions to $j=0$ with $W_q(0)=0$ 
for $q=\{+1,0,-1\}$. 

Note that $W_+(j) + W_0(j) + W_-(j) = j$,  which enables us to 
eliminate any one of the cumulative distributions, and rewrite in terms 
of the other two, for example, $W_0(j) = j - W_+(j)-W_-(j)$. 
If we denote a general sign $0,\pm1$ by $q$, then each of the 
$W_q(j)$ is an increasing function of $j$. Furthermore, we have 
\beq
W_+(N) = N_+, \qquad W_0(N)=N_0 , \qquad W_-(N) = N_- , 
\eeq
since  $N_+,N_0,N_-$ are the total number of $+$, $0$, $-$ 
states in the sample of $N = N_+ + N_0 + N_-$ individuals. 
In later analysis, we make use of the difference of these cumulative 
distribution functions $W_q(j)$, defined by  
\beq
w_q(j) = W_q(j) - W_q(j-1) . \label{wq-def}
\eeq
Intuitively, this is an appealing quantity to consider, since it represents 
the probability of site $j$ being occupied by an individual of genetic state $q$; 
however, in practice, the quantities $w_q(j)$ are either zero or one, 
depending which genetic state actually occurs in the data.  
In cases where a gene has an effect on phenotype, we expect $W_q(j)$ 
to be slowly varying in $j$, and so $w_q(j)$ could be obtained by 
taking averages over a range of neighbouring $j$ values. 

%---------------------
\subsection{{\bf Comparison of actual configuration with random allocation}
\label{comp-sec}}

In the random configuration, we assume that there is no correlation between 
phenotype and   we define the probabilities of $+1,0,-1$ states occuping 
any particular position in the list by  $w_+^{(0)}$, $w_0^{(0)}$, $w_-^{(0)}$, 
which are defined by  the probability density functions 
\beq
w_+^{(0)} = \frac{N_+}{N} , \qquad 
w_0^{(0)} = \frac{N_0}{N} , \qquad 
w_-^{(0)} = \frac{N_-}{N} . \label{wo-def}
\eeq
We use zero superscripts to denote the random configuration.  Note that 
$w_+^{(0)} + w_0^{(0)} + w_-^{(0)} = 1$ since each position in the list must 
be occupied. For this random state, the cumulative distributions for each 
genotype are given by summing (\ref{wo-def}) 
\beq
W_+^{(0)}(j) = j w_+^{(0)} , \quad 
W_0^{(0)}(j) = j w_0^{(0)} , \quad  
W_-^{(0)}(j) = j w_-^{(0)} , \label{WW0-def}
\eeq 
and note that these hold for $0 \leq j \leq N$. 

We now consider the difference between the actual cumulative distribution 
(\ref{WWq-def}) and the expected form for the random case (\ref{WW0-def}), 
\begin{align} 
\theta_+(j) =& \; W_+(j) - W_+^{(0)}(j) = W_+(j)  - \ds\frac{jN_+}{N} , \nn \\
\theta_-(j) = & \; W_-(j) - W_-^{(0)}(j) = W_-(j) - \ds\frac{jN_-}{N} . 
\label{Th-def}
\end{align}
As noted earlier, we do not need to consider the quantity 
$\theta_0(j)=W_0(j)-W_0^{(0)}(j)$, as any corresponding results 
can be obtained by noting that, for all $j$, we have 
$\theta_0(j) = -\theta_+(j)-\theta_-(j)$. 

The interpretation of the $\theta_q$-paths is that they describe the magnitude 
of the difference between the actual locations of individuals in the list 
and those expected from an average random allocation which would be 
given by $w_\pm^{(0)}$.

There are various properties of these $\theta_\pm(j)$ paths that are worth noting: 
\begin{itemize}
\item $\theta_+(0) = 0$, \ $\theta_-(0) = 0$, 
\item $\theta_+(N) = 0$, \ $\theta_-(N) = 0 $, this follows from 
	$W_q(N)=N_q$ and $W_q^{(0)}(N) = N_q$ (for $q=\{+1,0,-1\}$); 
\item $\theta\pm(j)\approx 0$ ($\forall j$) if there is no genotype-phenotype 
influence or correlation, since in this case, the expected distribution 
for the ordered state is the random configuration, and so any deviation 
from $(\theta_+(j),\theta_-(j)) = (0,0)$ will be due to random fluctuations.  
\end{itemize}
Hence the magnitude of $\theta_\pm(j)$ determines the 
strength of the effect of the genotype on the phenotype. 
We view the sequence of points $(\theta_+(j),\theta_-(j))$
as a path in two-dimensional space, which starts at $(0,0)$ at $j=0$, 
ends at $(0,0)$ when $j=N$, and makes some excursion away from $(0,0)$ 
for intermediate points $0<j<N$. 

To obtain the extremal $\theta_+$ values, let us consider the case where 
all $N_+$ occurrences of the $+$ states are in locations $1,2,\ldots,N_+$; 
The most extreme $\theta_+$ values is obtained by considering the case where 
all $N_+$ occurrences of the $+1$ states in locations $j=1,2,\ldots,N_+$, 
and the remaining items in the list are occupied by $0,-1$. This gives 
\begin{align}
W_+(N_+) = & N_+ , & \quad 
W_0(N_+) = & 0 , &  \quad 
W_-(N_+) = & 0 , \nn\\ 
W_+^{(0)}(N_+) = & N_+w_+^{(0)} , & \quad 
W_0^{(0)}(N_+) = & N_+ w_0^{(0)}, &  \quad 
W_-^{(0)}(N_+) = & N_+ w_-^{(0)} , \nn\\
\theta_+(N_+) = & N_+ - N_+w_+^{(0)} , & \quad 
\theta_0(N_+) = & -N_+ w_0^{(0)} , &  \quad 
\theta_-(N_+) = & -N_+ w_-^{(0)} . \nn\\ &&&&& 
\end{align}
Since $N = N_+ + N_0 + N_-$, we note that 
\beq
\theta_{+max} = \theta_+(N_+) = N_+(1-w_+^{(0)}) = \frac{N_+}{N} ( N- N_+ ) , 
= \frac{N_+(N_0+N_-)}{N} , 
\eeq
Similar calculations for the $0,-1$ gene states give 
\begin{align}
\theta_{-max} = \frac{N_-(N-N_-)}{N} = \frac{N_-(N_0+N_+)}{N} , \quad &
\theta_{0max} = \frac{N_0(N-N_0)}{N} = \frac{N_0(N_++N_-)}{N} . 
\end{align}

\begin{table} % TABLE 1 
\caption{Summary of variables/parameters in the model}
\label{var-tab} 
\begin{tabular}{lll}
\hline\noalign{\smallskip}
Variable/Parameter &  & Description \\
\noalign{\smallskip}\hline\noalign{\smallskip}
$N$ &  & Total number of individuals in sample \\
$N_+,N_0,N_-$ & & Number of individuals of each genotype \\
$W_+(j),W_0(j),W_-(j)$ & & Cumulative distribution of genotypes 
(\ref{WWq-def-th}) \\ 
$w_+(j),w_0(j),w_-(j)$ & & Probability density of genotypes \\
$\theta_+(j),\theta_-(j)$ & & Difference between cdf of 
actual and random configurations\\
\noalign{\smallskip}\hline\noalign{\smallskip}
$C_*(w_q)$ & & Constraints on the system, (\ref{Con1}), (\ref{Con2}) \\
$\alpha_*$ & & Lagrange multipliers -- used to solve the constrained problem \\
$S[w_q]$ & & Informational Entropy (Shannon Information) given by (\ref{Sdef}) \\
$u_q(j)$ & & Genotype field ($1 \leq j \leq N$, $q=\{+1,0,-1\}$)  \\
$E[w_q]) $ & & Genotype-phenotype interaction term (\ref{Edef}) \\
$\mathcal{A}[w_q , \alpha_*]$ & & Informational Action (\ref{ActionA})  \\
\noalign{\smallskip}\hline
\end{tabular}
\end{table}

%--------------------------------------------------
\section{{\bf Statistical Significance of $\theta-$paths} \label{pval-sec}}
\setcounter{equation}{0}

We noted in Section \ref{deriv-sec}, particularly in subsections \ref{comp-sec} 
that large deviations in the $\theta$-path away from zero correspond to 
highly significant genotype-phenotype interactions,  whilst $\theta-$paths 
which remain near $\theta=0$ for all $j$ are a sign of SNPs or genes that 
have less or no effect on phenotype.   \qb{A more rigorous theoretical basis 
for these effects will be given at the end of Section} \ref{math-sec}. 
% {\bf 6.13 rewritten to clarify reference foward } 

In this section we quantify the effect of observed genotype on observed 
phenotype by showing how to calculate the $p$-values for a given 
$\theta$-path, that is, trajectory given by $\theta_\pm(j)$ for $1 \leq j \leq N$. 
We do this by using the ideas of denisty of states from theoretical 
physics \cite{kittel}, where one considers the number different states 
for each energy level, and constructs a function which counts the number 
of states with energy below any particular certain energy.  
Here, we consider the number of possible paths that give rise 
to a deviation of $\theta(j)$ (or more) away from $\theta(j)=0$. 

We start with a simple system in which there are only two genetic 
states, $+1$ and $-1$, and later generalise to the three-state system 
($+1,0,-1$), in Section \ref{3p-sec}.

%---------------------------------------------------------
\subsection{{\bf Two-state significance calculation } \label{p2-sec}}

We assume that there are a given numbers, $N_+$ and $N_-$, of the 
$+1$ and $-1$ genetic states, and $N=N_++N_-$ (so there are $N_0=0$ 
zero genotypes).  We assume both $N_+,N_-$ are large, so that  factorials 
can be approximated using Stirling's formula ($N! \approx N^N \ee^{-N} 
\sqrt{2\pi N}$) \cite{olver}; to simplify notation in later calculations, we write 
$w=w_+^{(0)}$, so that $N_+ = Nw$ and $N_-=N(1-w)$.  The total number 
of $\theta$-paths is given by the number of ways that $+1$ and $-1$ can 
be ordered in a list, that is 
\beq
N_{tot} = \genfrac{(}{)}{0pt}{0}{N}{N_+} = \frac{ N! }{ N_+! \; N_-! } 
\sim \frac{ \ee^{-N [ w\log w + (1-w)\log (1-w) ]} }{\sqrt{2\pi N w (1-w)}} . 
\eeq

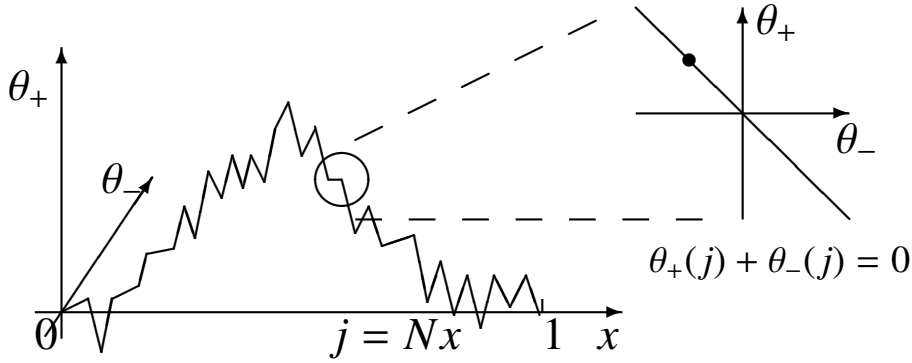
\begin{figure}[ht] % FIG 1 
\begin{picture}(400,150)(-50,-20)
\thicklines
% Main axes
\put(-10,0){\vector(1,0){220}}
\put(0,-10){\vector(0,1){110}}
\put(200,-14){{\Large $x$}}
\put(100,-14){{\Large $j = Nx$}}
\put(-20,80){{\Large $\theta_+$}}
\put(15,45){{\Large $\theta_-$}}
\put(-6,-9){\vector(2,3){40}}
\put(-10,-14){{\Large $0$}}
\put(180,-14){{\Large $1$}}
% path 
\put(0,0){\line(2,1){10}}
\put(10,5){\line(1,-4){5}}
\put(15,-15){\line(1,5){4}}
\put(19,5){\line(2,1){10}}
\put(29,10){\line(1,4){3}} 
\put(32,22){\line(5,1){10}}
\put(42,24){\line(1,4){4}}
\put(46,40){\line(1,-3){4}}
\put(50,28){\line(1,5){5}}
\put(55,53){\line(1,-2){5}}
\put(60,43){\line(1,4){4}}
\put(64,59){\line(1,-3){4}}
\put(68,47){\line(1,4){3}}
\put(71,59){\line(1,-2){5}}
\put(76,49){\line(1,5){4}}
\put(80,69){\line(1,2){5}}
\put(85,79){\line(1,-4){5}}
\put(90,59){\line(1,2){5}}
\put(95,70){\line(1,-4){5}}
\put(100,50){\line(1,0){5}} %midpoint of circle 
\put(105,50){\line(1,-4){5}}
\put(110,30){\line(1,2){5}}
\put(115,40){\line(1,-3){5}}
\put(120,25){\line(3,1){12}}
\put(132,29){\line(1,-5){5}}
\put(137,4){\line(1,3){5}}
\put(142,19){\line(1,-4){5}}
\put(147,-1){\line(1,3){5}}
\put(152,14){\line(1,-4){5}}
\put(157,-6){\line(1,4){5}}
\put(162,14){\line(1,-2){6}}
\put(168,2){\line(1,2){6}}
\put(174,14){\line(1,-3){5}}
\put(180,0){\line(0,1){5}}
% blowup (center 255,75)
\put(105 , 50){\circle{20}}
\multiput(110 , 65)(20,10){5}{\line(2,1){10}}
\multiput(110 , 35)(20,0){7}{\line(1,0){10}}
\put(215 ,  75){\vector(1,0){80}}
\put(255 ,  35){\vector(0,1){80}}
\put(215 , 115){\line(1,-1){80}}
\put(290 ,  60){{\Large $\theta_-$}}
\put(260 , 105){{\Large $\theta_+$}}
\put(235 ,  95){\circle*{5}}
\put(220,15){{\large $\theta_+(j)+\theta_-(j)=0$}}
% end
\end{picture}
\caption{Illustration of an example trajectory for the case of two genetic 
states; the three-dimensonal trajectory $(j,\theta_+,\theta_-)$,  
can be viewed, for any fixed $j$ (with $1 \leq j \leq N$), as a point 
in two-dimensional space $\btheta(j) = (\theta_+(j),\theta_-(j))$. 
However, the point is not free to arbirarily in the plane, it 
as to start and end at $(0,0)$, (at $j=0,N$) and in between, 
it is constrained to move on the line $\theta_- = - \theta_+$. }
\label{2d-th-fig}
\end{figure}

The cumulative distribution function $W_+(j)$ (\ref{WWq-def}) is the total 
number of $+$ states in the first $j$ elements of the list, and $W_+^{(0)}(j)$ %7-12
\qb{is} the expected number of $+$ states if the listing was random (\ref{WW0-def}), 
that is $W_+^{(0)}(j) = j w_+^{(0)}$.  The $\theta$-path is defined by 
$\theta_+(j) = W_+(j) - W_+^{(0)}(j) = W_+(j) - j w_+^{(0)}$ %7-10 
(\ref{Th-def}).    If there are $k$ `+1' states and 
so $(j-k)$ `-1' states in the first $j$ list positions ($1,2,\ldots,j$),  then we have 
\beq
\theta_+(j) = k - j w_+^{(0)}, \label{thetajk}
\eeq
and the number of ways that this can happen is 
\beq
N_{+paths}(k,j) = \genfrac{(}{)}{0pt}{0}{ \; j \; }{ \; k \; } \; 
\genfrac{(}{)}{0pt}{0}{N-j \; }{N_+-k\; } 
= \frac{ j! \; (N-j)! }{  k! \; (j-k)! \; (N_+-k)! \; (N-j-N_++k)! } . 
\eeq
This is the number of ways of allocating $k$ copies of the $+1$ state in the 
first $j$ locations multiplied by the number of ways of allocating $N_+-k$ 
copies of the $+1$ state in the last $N-j$ positions.    Whilst, formally we 
have $\theta_+(j)$ and $\theta_-(j)$, it is sufficient for us to consider only 
one of them, since  $\theta_-(j) = - \theta_+(j)$.  
In the two-dimensional space $(\theta_+,\theta_-)$, this can be viewed 
as motion in $j$ being constrained to the line $\theta_+(j) + \theta_-(j)=0$, 
as illustrated in Figure \ref{2d-th-fig}. %7.1
We assume \qb{$\theta_+(j)>0$} and then, to calculate a $p$-value, we want 
to know what fraction of all possible paths ($N_{tot}$), have a $\theta_+(j)$ 
value which is \qb{larger}  than (\ref{thetajk}). 
% {\bf 8.2 explain more extreme than (3.2) } % 8.2
\qb{Larger values of $|\theta_q(j)|$ correspond stronger dependencies 
of phenotype on genotype, which are less likely to occur by chance. 
We view such occurrences as being more `extreme', 
and wish to include all values of $\theta$ above $\theta_+(j)$ when 
determining a probability of an event of $\theta_+(j)$ occurring}. 
Thus we wish to evaluate 
\beq
p(\tilde k,j) =  \frac{1}{N_{tot}}  \sum_{k =\tilde k}^\infty N_{+paths}(k,j) . 
\label{p-sum1} \eeq

\begin{figure} % FIG 2 
\includegraphics[scale=0.5]{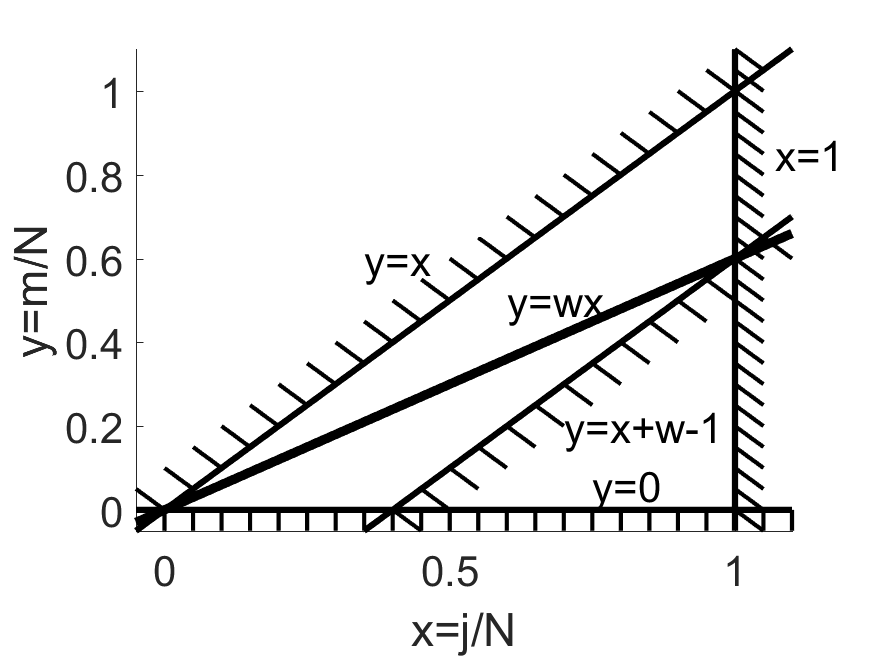}
\caption{Illustration of the region of interest in $(x,y)$-space, namely that 
satisfying all the constraints.  The thicker line shows the location of 
the maximum over $y$ for any fixed value of $x$.  
The constraints are $0<y<x<1$ and $y>x+w-1$, corresponding to 
$0<k < j < N$  (number of $+$ states,$k$, cannot exceed location, 
$j$, and both must be between zero and $N$) and $k >j+N_+ - N$, 
which is equivalent to $N- j > N_+-k$, so that there must be more 
positions in the list ($j+1,\ldots,N$) remaining than $+$ states still 
to allocate ($N_+-k$). }
\label{region-fig2}
\end{figure}

In the following calculations, we assume 
\beq
N \gg 1 , \quad N_+ = N w, \quad j = N x , \quad 
k = Ny , \quad \tilde k = N z , \label{Nassump} 
\eeq
so, for large lists, we expect $j,k$ to be relatively large too, 
% and $x,y,w=\mathcal{O}(1)$;  
hence 
\begin{align}
N_{+paths}(k,j) \sim & \; \frac{   j^j \, ( 1-j/N)^{N-j}  \sqrt{ j \, (N-j) } } 
{2\pi \,N \, k^k \, (j-k)^{j-k} \, (w-k/N)^{Nw-k} (1-w-j/N+k/N)^{N-Nw-j+k} 
\sqrt{R} } , \nn\\
R = & \; k (j-k) (w-k/N) (1-w-j/N+k/N) . 
\end{align}
The relative position in the list is then given by \qb{$0 < x < 1$}, 
and \qb{$0 < y < x$}.  Since the terms inside the square roots % 8.6
need to be positive, we also have \qb{$w+x-1 < y < w$}. 
The domain of interest is illustrated in Figure \ref{region-fig2}.

To evaluate (\ref{p-sum1}) we approximate by considering $x,y,w$ as 
continuous variables, and replacing the sum (\ref{p-sum1}) by the integral  
\begin{align}
\tilde p(z,x,w) = & \; \frac{\ee^{ N g(w,x) }\sqrt{w(1-w)x(1-x) } } {\sqrt{2\pi N}} 
\int_{y = z}^1 \frac{ \ee^{-N f(y,w,x) } }{ \sqrt{y(x-y)(w-y)(1+y-w-x) } }  \, N \, \dd y 
\nn \\ 
f(y,w,x) = & \; y\log y + (x-y) \log (x-y) + (w-y)\log(w-y) \nn \\ 
     & + (1+y-x-w)\log(1+y-w-x) , \nn \\ 
g(w,x) = & \; w\log w + (1-w)\log(1-w) +x\log x + (1-x) \log(1-x) . 
\end{align} 
The dominant part of this integral comes from minimum of $f(y,w,x)$ over 
$y$, which is given by $y=wx$ (from solving $f_y=0$ for $y$).  Since 
\beq 
f_{y} (y,w,x) = \log \left( \frac{y(1+y-x-w)}{(x-y)(w-y)} \right) , \quad \mbox{and} \quad 
\left. f_{yy} (y,w,x) \right|_{y=wx} = \frac{1}{xw(1-x)(1-w)} , 
\eeq
%and $f(y,x,w)|_{y=xw} = g(x,w)$,  % 9.4 
we have 
\begin{align}
\tilde p(z,x,w)  = &\; \frac{ \sqrt{N} }{ \sqrt{2\pi} \sqrt{ w(1-w)x(1-x) }} \int_{y=z}^1 
\exp \left(- \frac{ N (y-xw)^2 }{ 2 x w (1-x) (1-w)} \right) \,  \dd y  \nn \\  = & \;  
\frac{1}{2} \erfc \left( \frac{\sqrt{N}(z-xw) }{\sqrt{2xw(1-x)(1-w)}} \right) . 
\nn \\ & \end{align}
Using (\ref{thetajk}) and (\ref{Nassump}), and noting that we should 
consider both tails of the distribution \qb{($\theta > \theta_+(j)$ and 
$\theta < - \theta_+(j)$)}, % {\bf explain $\theta_c$} % 9.7 
we double this value of $p$, giving  
\beq 
p_+(j) = \erfc \left( \frac{ \left| \theta_+(j) \right|}{\sqrt{2Nxw(1-x)(1-w)}} \right) 
 = \erfc \left( \frac{ N\sqrt{N} \; \left| \theta_+(j) \right|}{\sqrt{2 j (N-j) N_+ N_- }} \right) . 
\label{p2final} \eeq 
(where $|z| = z$ if $z\geq 0$ and $|z| = -z$ if $z<0$). 

This formula gives a $p$-value for each position in the list, $j$; 
however, it would be preferable to have a single $p$-value for 
each SNP, thus we now propose various formula for 
obtaining a single $p$-value from the whole list, (\ref{p2final}). 
Firstly, we could simply take the minimum over all $j$-values 
\beq  
p_{\mbox{{\scriptsize SNP1}}} = \min_{1 \leq j \leq N} \{  p_+(j) \}  , 
\label{psnpmin}
\eeq
or we could consider the average (mean) $p$-value calculated over every 
position in the list 
\beq
p_{\mbox{{\scriptsize SNP2}}} = \frac{1}{N} \sum_{j=1}^N p_+(j) . 
\label{psnp2}
\eeq
Since we commonly want to know the outliers, and so plot 
$L = -\log p_{\mbox{{\scriptsize SNP}}}$, one could also plot 
\beq 
L_{\mbox{{\scriptsize SNP3}}} = \frac{1}{N} \sum_{j=1}^N -\log p_+(j) . 
\label{psnp3}
\eeq
Our final two methods rely on taking various weighted averages of 
$|\theta_q(j)|$ or $\theta_q(j)$ over $j$.   Since (\ref{p2final}) can be 
written as 
\beq
p_+(j) = \erfc(Z), \quad \mbox{with} \;\;\;  
Z = \frac{ |\theta_+(j)| N \sqrt{N} }{ \sqrt{2j(N-j) N_+N_-}} , 
\eeq
we consider 
\beq
p_{\mbox{{\scriptsize SNP4}}} = \erfc(Z), \quad 
Z = \frac{\sqrt{N}}{\sqrt{2N_+N_-}} \sum_{j=1}^{N} \frac{ |\theta_+(j)|}{\sqrt{j(N-j)}} , 
\label{psnp4}
\eeq
and
\beq
p_{\mbox{{\scriptsize SNP5}}} = \erfc(|Z|), \quad 
Z = \frac{\sqrt{N}}{\sqrt{2N_+N_-}} \sum_{j=1}^{N} \frac{ \theta_+(j)}{\sqrt{j(N-j)}} . 
\label{psnp5}
\eeq
The efficacy of these will be considered in Section \ref{num-sec}.

%---------------------------------------------------------
\subsection{{\bf Three-state significance calculation } \label{3p-sec}}

We now consider the case of 3 genetic states, $+1,0,-1$ and aim to 
determine a formula similar to (\ref{p2final}) for the $p$-value in this 
three-component case.  
In the 2-genetic-state system, if we assume only $+1$ and $-1$ genetic 
states occur, the distance of $\mathbf{\theta} = (\theta_+,\theta_-)$ from 
$(0,0)$ is given by $d = \sqrt{\theta_+^2 + \theta_-^2} = \sqrt{2\theta_+^2} 
= \sqrt{2} \, |\theta_+|$,  so using $|\mathbf{\theta}| = |\theta_+| = |\theta_-|$ 
is consistent with $|\mathbf{\theta}|=d$, the difference being only a factor 
of $\sqrt{2}$, and calculations of $p$-values based on the density of states 
is not changed by how we calculate $|\mathbf{\theta}|$. 
With two states, the two-dimensional motion on the $(\theta_+,\theta_-)$ 
plane is contrained to the line $\theta_++\theta_-=0$, as illustrated in 
Figure \ref{2d-th-fig}.

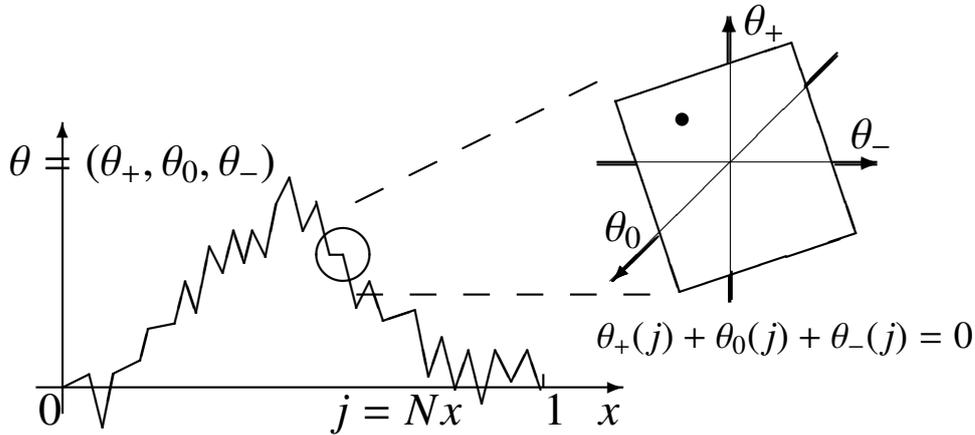
\begin{figure}[!ht] % FIG 3 
\begin{picture}(400,170)(-50,-20)
\thicklines
\put(-10,0){\vector(1,0){220}}
\put(0,-10){\vector(0,1){110}}
\put(200,-14){{\Large $x$}}
\put(100,-14){{\Large $j = Nx$}}
\put(-20,80){{\Large $\mathbf{\theta} = (\theta_+,\theta_0,\theta_-)$}}
\put(-9,-14){{\Large $0$}}
\put(180,-14){{\Large $1$}}
% path 
\put(0,0){\line(2,1){10}}
\put(10,5){\line(1,-4){5}}
\put(15,-15){\line(1,5){4}}
\put(19,5){\line(2,1){10}}
\put(29,10){\line(1,4){3}}
\put(32,22){\line(5,1){10}}
\put(42,24){\line(1,4){4}}
\put(46,40){\line(1,-3){4}}
\put(50,28){\line(1,5){5}}
\put(55,53){\line(1,-2){5}}
\put(60,43){\line(1,4){4}}
\put(64,59){\line(1,-3){4}}
\put(68,47){\line(1,4){3}}
\put(71,59){\line(1,-2){5}}
\put(76,49){\line(1,5){4}}
\put(80,69){\line(1,2){5}}
\put(85,79){\line(1,-4){5}}
\put(90,59){\line(1,2){5}}
\put(95,70){\line(1,-4){5}}
\put(100,50){\line(1,0){5}} %midpoint of circle 
\put(105,50){\line(1,-4){5}}
\put(110,30){\line(1,2){5}}
\put(115,40){\line(1,-3){5}}
\put(120,25){\line(3,1){12}}
\put(132,29){\line(1,-5){5}}
\put(137,4){\line(1,3){5}}
\put(142,19){\line(1,-4){5}}
\put(147,-1){\line(1,3){5}}
\put(152,14){\line(1,-4){5}}
\put(157,-6){\line(1,4){5}}
\put(162,14){\line(1,-2){6}}
\put(168,2){\line(1,2){6}}
\put(174,14){\line(1,-3){5}}
\put(180,0){\line(0,1){5}}
% blowup (center 250,85)
\put(105 , 50){\circle{20}}
\multiput(110 , 70)(20,10){5}{\line(2,1){10}}
\multiput(110 , 35)(20,0){6}{\line(1,0){10}}
\thinlines
\put(200, 85){\line(1,0){100}}
\put(250, 37){\line(0,1){90}}
\put(290,125){\line(-1,-1){85}}
\thicklines
\put(200,85){\line(1,0){15}}
\put(200,84){\line(1,0){15}}
\put(249.5, 32){\line(0,1){10}}
\put(250, 32){\line(0,1){10}}
\put(250.5, 33){\line(0,1){11}}
\put(290,126){\line(-1,-1){12}}
\put(290,125){\line(-1,-1){12}}

\put(288, 85){\vector(1,0){17}}
\put(289, 84){\vector(1,0){16}}
\put(250,123){\vector(0,1){17}}
\put(249,122){\vector(0,1){18}}
\put(223,57){\vector(-1,-1){17}}
\put(222,56){\vector(-1,-1){16}}
\put(221,55){\vector(-1,-1){15}}

\put(295, 90){{\Large $\theta_-$}}
\put(255,133){{\Large $\theta_+$}}
\put(203, 55){{\Large $\theta_0$}}
\put(232,101){\circle*{5}}
\put(207,108){\line(3,1){66}}
\put(207,108){\line(1,-3){24}}
\put(273,130){\line(1,-3){24}}
\put(231, 36){\line(3,1){66}}
\put(200,15){{\large $\theta_+(j)+\theta_0(j)+\theta_-(j)=0$}}

\end{picture}
\caption{Illustration of an example trajectory for the case of three genetic 
states.  Here,  the $\btheta$-path is can be viewed as a four-dimensional 
object ($j,\theta_+(j),\theta_0(j),\theta_-(j)$) with $1 \leq j \leq N$;  or as a 
three-dimensional trajectory in $\btheta = (\theta_+(j),\theta_0(j),\theta_-(j))$ 
space, which starts and ends at $\btheta=(0,0,0)$; however, this 
trajectory is constrained to lie on the plane $\theta_+ + \theta_0 + \theta_-=0$.  }
\label{3d-th-fig}
\end{figure}

However, in a system with three genetic states, it is not immediately 
clear how best to interpret the distance of $\mathbf{\theta}$ from zero.  
Arbitrarily choosing $|\mathbf{\theta}|$ as $|\theta_+| + |\theta_-|$ or 
$\sqrt{\theta_+^2+\theta_-^2}$ ignores the role that $\theta_0 = 
-\theta_+ - \theta_-$ has in making the $\theta-$paths move away from zero. 
We consider the full 3D motion of  $\mathbf{\theta} = 
(\theta_+ , \theta_0, \theta_-)$ as illustrated in Figure \ref{3d-th-fig}, 
which illustrates a trajectory (or $\theta-$path) which starts from $j=1$ 
(corresponding to $x=0$) and ends at  $j=N$ (corresponding to $x=1$).  
At each point (labelled by $j$ or $x$)  along the path, we have values for 
$\mathbf{\theta}=(\theta_+,\theta_-,\theta_0)$. 
At any particular location $j$, we treat the three $\theta_q(j)$ variables 
in a consistent manner.   We calculate the distance from the origin 
$(0,0,0)$ to $\mathbf{\theta}$, and then make use of the condition 
that the trajectory is constrained to lie on the plane $\theta_+ + 
\theta_0+\theta_-=0$ afterwards, which yields the distance $d$ as 
\beq
d^2 = \theta_+^2 + \theta_-^2 + \theta_0^2 
= 2 \theta_+^2 + 2 \theta_-^2 + 2 \theta_+\theta_- 
= \half \left[ 3 (\theta_+ + \theta_-)^2 + (\theta_+ - \theta_-)^2 \right]  . 
\eeq

We consider the case where, in locations $1,2,\ldots,j$ in the ordered list, 
there are $k$ occurrences of the $+1$ genetic state, and $l$ occurrences 
of $-1$. Thus 
\begin{align}
W_+(j) = k , \quad & W_+^{(0)}(j) = j w_+^{(0)} , \quad & \theta_+(j) = k - j w , 
\quad & w =w_+^{(0)}  \nn \\ 
W_-(j) = l , \quad & W_-^{(0)}(j) = j w_-^{(0)} , \quad & \theta_-(j) = l - jv , 
\quad & v =w_-^{(0)} .  
\label{3cpt-Wtheta} \end{align}

As in section \ref{p2-sec} we assume 
\begin{align}
j = N x , \quad k = N y, \quad l = N z , \quad 
N_+ = N w, \quad N_- = N v , \quad N \gg1, 
\label{3cpt-scales}  \end{align}
so that there are many occurrences of each genotype, 
and we consider the main central part of the trajectory (that is, 
$j$ is not near $j=1$ or $j=N$), so there are many of each genetic 
state in the intervals $1\ldots j$ and $j\ldots N$. 
This assumption simplifies later calculations by allowing Stirling's 
formula to be used \cite{olver}. 

We calculate the total number of paths which have $N_+,N_0,N_-$ 
occurrences the genetic states $+1,0,-1$ respectively, as  
\beq
N_{\mbox{{\footnotesize{tot}}}} = \frac{N!}{N_+! \; N_0! \; N_-\!} 
\approx \frac{\ee^{-N[g(w) + g(v) - g(1-v-w) ]}}{2\pi N} , \qquad 
g(q) = q\log(q) , 
\eeq
by Stirling's formula \cite{olver}, and the number of paths from 
$(j,\theta_+,\theta_-)=(0,0,0)$ to $(j,k-jw,l-jw)$ and on to (N,0,0), as  
\begin{align}
N_{\mbox{{\footnotesize{paths}}}}(j,k,l) = & \frac{ j! }{ k! \; l! \; (j-k-l)! } . 
\frac{(N-j)!}{ (N_+-k)! \; (N_- - l)! \; (N-N_+-N_- - j + k + l)! } . 
\nn \\ & 
\end{align}
The conditions of there being a positive number of each state in 
both the intervals $\{ 0, \ldots , j \}$ and $\{ j , \ldots , N \}$ imply that 
\beq % 3.22
\mbox{\qb{$0 <  y <  x <  1,\qquad 0 < z < x <  1,\qquad x+v+w-1 <  y+z < x .$}}
\eeq
The last inequalities arise from the fact that there must be 
$j-k-l = N(x-y-z) > 0$ occurrences of the zero state in the first $j$ elements, 
and that there must be $N-N_+-N_- - j + k + l = N(1-w-v-x+y+z) > 0$ 
zero states in the last $N-j$ elements. 

We define $\widehat p(j,k,l)$ as the fraction of all possible paths, 
that have $k$ occurrences of the +1 genetic states, and $l$ occurrences 
of the state $-1$ in the first $j$ items of the list.  This is given by the path 
density $\widehat p(j,k,l) = N_{\mbox{\footnotesize{paths}}}(j,k,l) / 
N_{\mbox{\footnotesize{tot}}}$ which can be approximated by 
\begin{align}
\widehat p(j,k,l) = & \; \frac{\ee^{N G(x,y,z) } }{2\pi N} \sqrt{ 
\frac{x (1-x) v w (1-v-w)}{ y z (w-y)(v-z)(x-y-z)(1+y+z-x-v-w) }} , \nn \\ 
G(x,y,z) = &\; g(x) + g(1-x) + g(w) - g(y) - g(w-y) 
+ g(v)  - g(z) - g(v-z) \nn \\ & + g(1-v-w) - g(x-y-z) - g(1+y+z-w-v-x)   
\label{phat-app1}
\end{align}

The function $G(x,y,z)$ has a single maximum, a property which can be 
demonstrated by solving the conditions $G_y=0=G_z$ for $y,z$, where 
\begin{align}
\pad{G}{y} =  \log \left( \frac{(w-y)(x-y-z)}{y(1-v-w-x+y+z)} \right) , & \quad 
\pad{G}{z} = \log \left( \frac{(v-z)(x-y-z)}{z(1-v-w-x+y+z)} \right) .  
\end{align}
This gives $y=xw$, $z=xv$. 
Evaluating the second derivatives at the stationary point gives
$H=G_{yy} G_{zz} - G_{yz}^2 > 0$,  and since $G_{yy}<0$, 
the stationary point at $y=xw$, $z=xv$ is a maximum.

\begin{figure} % FIG 4 
\hspace*{-19mm}
\includegraphics[scale=0.65]{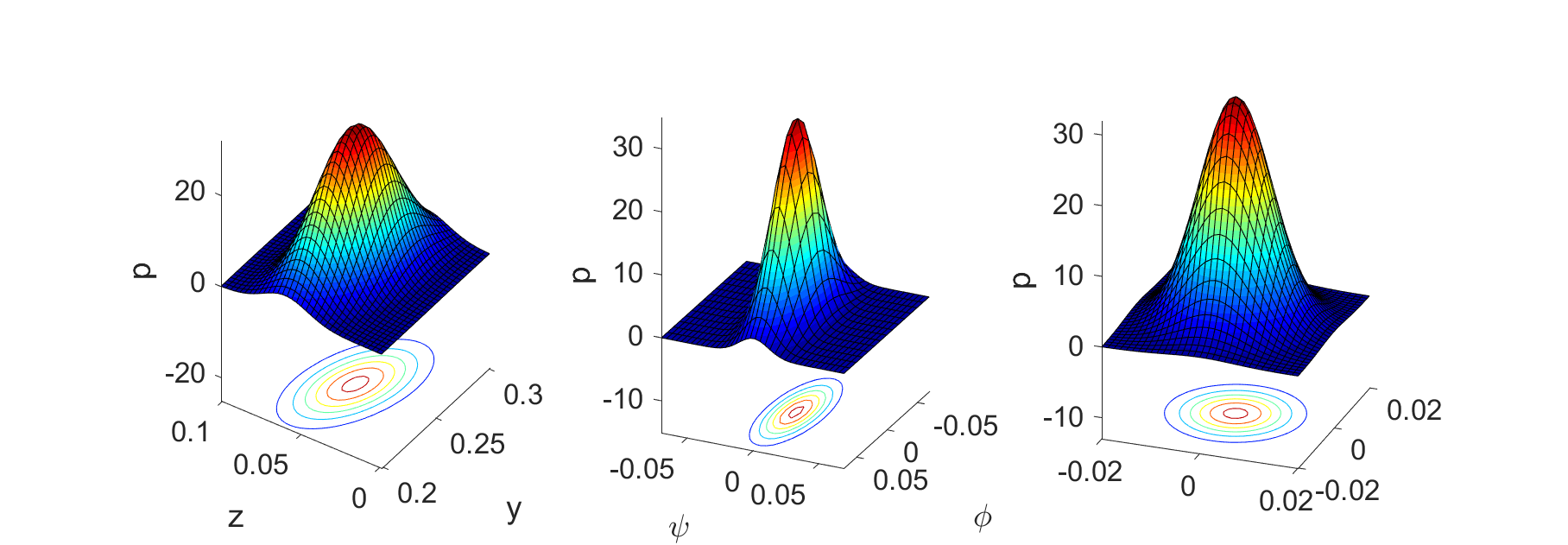}
\caption{Illustration of the two-dimensional distribution of 
path densities: 
left - as a function of $(y,z)$ in which the distribution exhibits 
non-zero covariance (\ref{Ghyz}); 
centre - as function of $(\phi,\psi)$ in which there is no correlation, 
but the variances differ (\ref{Gpp}); 
right - after the transformation to $(\ro,\eta)$, given by (\ref{Gro}) 
\qb{in which the distribution is cylindrically symmetric -- being 
only a function of $\ro$}. }
\label{dist2d-fig}
\end{figure}

Since $\left. G(x,y,z) \right|_{y=xw,z=xv} = 0$, we can approximate the 
dominant term in the number of paths formula (\ref{phat-app1}) as 
\begin{align}
\widehat p(j,k,l) \sim & \;  \frac{1}{ 2 \pi N x (1-x) v w (1-w-v) } \exp 
\left( - \frac{ N \widehat G(y,z) }{ 2 v w x (1-x) (1-v-w) }\right) , \nn \\ 
\widehat G(y,z) = &\; v(1-v) (y-wx)^2 + w(1-w) (z-vx)^2 + 2 v w (y-wx)(z-vx) . 
\label{Ghyz}
\end{align}
The transformation $y = wx + \phi + w \psi$,  $z = vx - \phi + v \psi$, 
equivalent to 
\begin{align}
\phi  = &\; \frac{vy - wz}{v+w} , \qquad \psi = \frac{ y-xw+z-xv }{v+w} , 
\label{phipsi}
\end{align}
`diagonalises' this system (\ref{Ghyz}) to 
\begin{align}
\widehat p(j,k,l) \sim & \;  \frac{1}{ 2 \pi N x (1\!-\!x) v w (1\!-\!w\!-\!v) } 
\exp \left( - \frac{ N (v\!+\!w) \, \left[ \, (1\!-\!v\!-\!w) \phi^2 + v w \psi^2 \,\right]  }
{ 2 v w x (1-x) (1-v-w) }\right) . \nn \\ \label{Gpp}
\end{align}
Figure \ref{dist2d-fig} shows how the transformation (\ref{phipsi}) removes 
the correlation present in the multi-dimensional distribution (\ref{Ghyz}). 
Note that in the left panel, the major and minor axes of the elliptic contours 
do not align with $y,z$ axes, whereas in the centre panel they do.  
Whilst (\ref{Gpp}) is simply the the product of two Gaussians, 
since their standard deviations differ, we propose a further 
transformation to a new variable in all points with the same distance 
from the origin have the same probability, as illustrated in the 
right-most panel of figure \ref{dist2d-fig}, which has circular contours. 
We define $\ro=\ro(\phi,\psi)$ by 
\beq
\ro^2=(1-v-w)\phi^2+vw\psi^2, \label{ro-def}
\eeq 
so that 
\beq
\psi = \frac{\ro \cos \eta}{\sqrt{vw}}, \qquad 
\phi = \frac{\ro \sin\eta}{\sqrt{1-v-w}} , \qquad 
\tan\eta = \frac{\phi}{\psi} \sqrt{\frac{1-v-w}{vw}} , 
\label{psiphi}
\eeq
yields the path density as 
\beq
\widehat p(j,k,l) \sim  \frac{1}{ 2 \pi N x (1\!-\!x) v w (1\!-\!w\!-\!v) } 
\exp \left( - \frac{ N (v\!+\!w) \, \ro^2 }{ 2 v w x (1-x) (1-v-w) } \right) . 
\label{Gro}
\eeq
To obtain a $p$-value for the number of paths with more extreme 
$\btheta$-variations, we have to integrate this quantity over the range of 
$k=Ny,l=Nz$, or equivalently $\ro$ values for which $\widehat p$ is smaller. 

To make this statement precise, we have to define what we mean 
by `more extreme' values of $\theta_\pm$, that is, 
%what combination 
%of $k,l$ (equivalently, $y,z$ or $\phi,\psi$, or $\ro$) qualify as more extreme 
\qb{we consider the range of all possible $\theta_+(j),\theta_-(j)$ values, 
rescale these via $\theta_+(j) =N(y-xw)$ and $\theta_-(j) = N(z-xv)$, 
using} (\ref{3cpt-Wtheta}) and (\ref{3cpt-scales}), \qb{and then perform 
the transformations}  (\ref{phipsi}) and (\ref{ro-def})  
\qb{to obtain the corresponding values $\phi,\psi$ and ultimately $\ro$. 
In terms of $\ro$ the resulting probability density function} (\ref{Gro}) 
\qb{is cylindrically symmetric, as illustrated in Figure} \ref{dist2d-fig}.  
\qb{Given a specific set of realised values for $\theta_+(j),\theta_-(j)$, 
we perform these same transformations and which give us a `threshhold' value, 
$\rho_c$, defined by } 
\begin{align}
\ro_c^2 = & \; \frac{ v(1-v) \theta_+^2 + w(1-w) \theta_-^2 + 2 v w \theta_+\theta_- }
{ N^2 (v+w) } , \nn \\ = & 
\frac{ N_-(N-N_-) \theta_+^2 + N_+(N-N_+) \theta_-^2 + 2 N_- N_+ \theta_+\theta_- }
{ N^3 (N_+ + N_-) } . 
\end{align}
%{\bf explain how $\rho_c$ defd and explain more fully } % 3.30 
\qb{To sum the probability density function $\widehat p(j,k,l)$ over all $\theta$-paths 
which have {\em{lower}} probabilities of occurring, we integrate $\widehat p(j,k,l)$ 
over the range $\ro_c < \ro < \infty$. }   \qb{Hence} we obtain 
\begin{align}
p = & \sum_k \sum_l  \hat p = N^2 \int\!\!\!\!\int_D \hat p \, \dd y \dd z 
= N^2 (v\!+\!w) \int\!\!\!\!\int_D \hat p \,\dd \phi \dd \psi 
= \frac{ N^2 (v+w) }{\sqrt{vw(1-v-w)}} \int \!\!\!\!\int_D \hat p \, 
\ro \, \dd\ro \dd\eta \nn \\ 
= & \frac{ 1 }{ 2 \sqrt{vw(1-v-w)} } \exp\left( - 
\frac{ N(v+w) \ro_c^2 }{ 2vwx(1-x)(1-v-w)) } \right) . 
\end{align}
Here, the domain of integration $D$ is given by all value of $(y,z)$ 
or equivalently $(\psi,\psi)$ which lead to a value of $\ro$ that is larger 
than $\ro_c$.  Any such path has a lower probability of occurring 
than the path observed.  

Inverting the trasnformations (\ref{phipsi})--(\ref{psiphi}), we obtain 
\begin{align}
p(\theta_+(j),\theta_-(j)) 
%= & \; \frac{ N\sqrt{N} }{ 2 \sqrt{ N_+ N_0 N_- } } 
% \exp \left( - \frac{ N^5 (N_+ + N_-)  \ro_c^2 }{ 2 j (N-j)  N_+ N_0 N_- } \right) \nn \\ 
= & \;\frac{ N\sqrt{N} }{ 2 \sqrt{ N_+ N_0 N_- } } 
\exp \left( - \frac{ N^2 \left[  N_-(N-N_-) \theta_+^2 + N_+(N-N_+) \theta_-^2 
+ 2 N_- N_+ \theta_+\theta_-  \right] }{ 2 j (N-j)  N_+ N_0 N_- } \right)  . 
\label{p3final}
\end{align} 
Note that this does {\em not} reduce to the two-state result (\ref{p2final}) 
in the limit of small $N_0$. 
As with (\ref{p2final})--(\ref{psnpmin}), equation (\ref{p3final}) 
gives a $p$-value for each position in the list, 
to give a value for the whole SNP, one could quote 
$\min_j \{ p(\theta_+(j),\theta_-(j)) \}$ as in (\ref{psnpmin}). 
Alternatively, either (\ref{psnp2})-(\ref{psnp3}) could be used,  or 
following (\ref{psnp4})-(\ref{psnp5}), we take an average value of 
the argument inside the exponential in (\ref{p3final}), and use 
\begin{align}
p_{\mbox{{\scriptsize SNP4}}} = & \; \frac{N \sqrt{N}}{2 \sqrt{N_+ N_0 N_-}} 
\exp \left( - \frac{N^2 Z }{ 2 N_+ N_0 N_- }  \right), \nn \\ 
Z = & \; \frac{ 1 }{ N } \sum_{j=1}^{N} \frac{ N_-(N-N_-)\theta_+(j)^2 + 
N_+(N-N_+)\theta_-(j)^2 + 2 N_+N_-\theta_+(j) \theta_-(j) }{ j (N-j) } . 
\end{align}

%----------------------------------------------------------------------------
\subsection{{\bf Summary}} \label{bio-p-sec}
% new discussion of significance of p-values 

\qb{ 
Typically, in an investigation of the effect of a particular genetic mutation 
or SNP, we would start with the `null hypothesis', which is a statistical 
assumption that there is no effect of the genetic state on the ordering 
of phenotypes observed in the population data. 
Thus one would expect the local distributions of the genetic states 
$+1,0,-1$ to be the same across the whole ordered list of phenotypes, 
and there would be only random fluctuations from the mean. This means 
that the $\theta$-path $\theta_\pm(j)$ would be small for all $j$ ($1\leq j\leq N$). 
Mathematically, we can write this as $w_q(j) \approx w_q^{(0)}$} 
(\ref{wq-def})--(\ref{wo-def}), 
\qb{which implies $\theta_\pm(j) \approx 0$} (\ref{Th-def}).  
\qb{If we were to observe a $\theta$-path which exhibits a `large' deviation from zero, 
we need to determine whether that could have occurred by chance, or whether 
it is `large' enough to be statistically significant, thus we would 
like to know the probability of it occurring under the null hypothesis. 
This is what the $p$-value tells us.   
One can choose whether to work at 5\% or 1\% threshold level;  
and if one is making multiple tests, a Bonferroni} \cite{bon} \qb{or 
Benjamini-Hochberg} \cite{bh-fdr} \qb{correction procedure can be applied. 

If the calculated $p$-value is smaller than the threshold value 
for a particular SNP, then there is evidence {\em against} the null hypotheis, 
and it is then reasonable to claim that for the SNP under consideration, 
genotype {\em has} an influence on phenotype. 
In the next sections, we analyse the form of this dependence, 
and provide a quantitative description of it. 
}

%----------------------------------------------------------------------------
\section{{\bf Mathematical model}} \label{math-sec}
\setcounter{equation}{0}

The mathematical model which provides an effective field strength 
that quantifies the genotype-phenotype interaction is derived from a 
combination of Shannon's information theory \cite{Shannon} and the 
Euler-Lagrange variational derivatives that are commonly used to 
derive the equations of motion in classical mechanics \cite{gold}. 

We now take a probabilistic approach to the $\theta$-paths (\ref{Th-def}) 
and allele distributions (\ref{WWq-def}), defining probability density functions 
\begin{align} 
w_+(j) =  \;W_+(j) - W_+(j\!-\!1) , \quad % \nn\\
w_0(j) =   \;W_0(j) - W_0(j\!-\!1) , \quad % \nn\\ 
w_-(j) =   \;W_-(j) - W_-(j\!-\!1) ,  \label{wq-def2}
\end{align}
which we interpret as the {\em probabilities} of finding each of gene state $\{+1,0,-1\}$ 
at site $j$ in the ordered state.  The probabilities $w_q(j)$ for $1 \leq j \leq N$ 
and $q=\{+1,0,-1\}$ are the basis of the probabilistic model. 
Note that in any particular set of observations each $w_q(j)$ is either 
zero or one, and the cumulative distributions $W_q(j)$ increase 
by zero or one as $j \mapsto j+1$.  In contrast, in the mathematical 
model we assume $W_q(\cdot)$ is a monotonically increasing function 
and we assume $w_q(j)$ vary slowly in $j$ so that they can be interpreted 
as a local (in $j$) probability of finding state $q$ at position $j$ in the list. 

Since one of the genetic states $q=\{+1,0,-1\}$ must be present at each 
site $j$, they must sum to one at each $j$, thus we have 
\beq
C_j := w_+(j) + w_0(j) + w_-(j) -1 = 0 , \quad (j=1,2,\ldots,N). 
\label{Con1}
\eeq
which is the first constraint on our system.  Summing each of (\ref{wq-def2}) 
over list positions, $j$, the cumulative distributions are given by 
\beq
W_+(j) = \sum_{i=1}^j w_+(i) , \qquad 
W_0(j) = \sum_{i=1}^j w_0(i) , \qquad 
W_-(j) = \sum_{i=1}^j w_-(i) , 
\label{WWq-def-th}  \eeq
The second constraint that we have to impose is that the total 
number of individuals with each genotype matches the data, that is 
\begin{align}  
C_+(w_+) := & \; \sum_{j=1}^N w_+(j) - N_+ =0 , \qquad 
C_0(w_0) := \sum_{j=1}^N w_0(j) - N_0 =0 , \nn \\  
C_-(w_-) := & \sum_{j=1}^N w_-(j) - N_- =0 .  
\label{Con2}  \end{align}

We propose to analyse the information content of the genetic states 
of the ordered arrangement, hence we introduce the Shannon entropy 
\begin{align}
S [ \bw ]  = & \; - \sum_{j=1}^N \left( w_+(j) \log w_+(j) + 
w_0(j) \log w_0(j) + w_-(j) \log w_-(j)  \right) .   \label{Sdef} 
\end{align}
Since each of the $w_q(j)$ variables is positive, the $\log$ terms are real 
and negative,  thus $S[\bw]$ is positive. The entropy $S[\bw]$ can only 
be zero in the case where, for every $j$ two of the $w_q(j)$ ($q\in\{+1,0,-1\}$) 
are zero and the other one equal to one.  Typically $S[\bw]$ is strictly positive, 
and we will show later that the maximum entropy occurs for the random 
configuration (\ref{wo-def}). 

In cases where the ordered state exhibits some correlation between 
phenotype and genotype we introduce `fields' to describe and help 
understand this effect. To preserve the equal treatment of the three states 
$\{+1,0,-1\}$,  we start with three fields, one for each genetic state, and each 
dependent on location, $\mathbf{u} =( u_+(j),u_0(j),u_-(j))$.  We propose 
a simple linear interaction term  relating the fields $\mathbf{u}$ to the location 
probabilities $\bw$ of the form 
\beq
E[\bw] =  - \sum_{j=1}^N \left( w_+(j) u_+(j) + w_0(j) u_0(j) + w_-(j) u_-(j) \right) . 
\label{Edef}  \eeq
%{\bf Explain energies  } % 4.6
The sign means that the minimum energy state is obtained when larger values 
of $w_q(j)$ coincide with larger values of $u_q(j)$.  We consider $E[\bw]$ to 
be similar to the potential energy in Lagrangian mechanics \cite{gold}. 
\qb{If we were to maximse entropy,} (\ref{Sdef}), \qb{then the genotypes 
would be randomly distributed across the ordered list of phenotypes; thus to 
account for some influence of the genetic state on the ordered phenotype list, 
we must include an extra factor (a genetic `force', `field', or `potential energy', 
denoted by $u_q(j)$) 
which can be interpreted as favouring a drift of a genetic state to higher 
or lower values of $j$, that is, towards one or other ends of the phenotype list}. 

In Lagrangian mechanics, ($\mathcal{L}$) is defined to be the difference 
of kinetic energy ($\mathcal{T}$) and potential energy ($\mathcal{V}$), and 
the action is the time integral of the Lagrangian, that is, $\mathcal{A}
 = \int \mathcal{L} \,\dd t = \int (\mathcal{T} - \mathcal{V}) \, \dd t$. 
The equations of motion are then obtained by taking variational derivative 
of the action with respect to path ($\bw$).  
In our information theory approach, we define action as 
$\mathcal{A}_1[\bw] = S[\bw] - E [\bw]$, and take the variational 
derivative with respect to the genetic state probabilities $\bw$. 
However, we are not free to consider all possible variations, 
we have to make sure that the constraints (\ref{Con1})--(\ref{Con2}) 
are satisfied, thus we use the method of Lagrange multipliers 
to include these contraints into the variational procedure. 

Combining the constraints (\ref{Con1}), (\ref{Con2}) with Lagrange 
multipliers $\balpha = (\alpha_1,\alpha_2,\ldots,\alpha_N)$, 
$\bbeta = (\beta_+,\beta_0,\beta_-)$ and the difference, $S - E$, 
we define the informational action $\mathcal{A}$, by 
\begin{align}
\mathcal{A} [ \bw , \balpha , \bbeta ] = & \; S[\bw] - E[\bw] + \beta_+ C_+(w_+) 
+ \beta_0 C_0(w_0) + \beta_- C_-(w_-)  + \sum_{j=1}^N \alpha_j C_j(\bw) . 
\label{ActionA} \end{align}
The location probabilities $\bw$ are then given by requiring the first variation 
of $\mathcal{A}(\bw,\balpha,\bbeta)$ with respect to $w_q(j)$ to be zero. 
The constraints are recovered and satisfied by requiring the first variation 
of $\mathcal{A}$ with respect to each element of $\balpha,\bbeta$ to be zero. 
The details of these calculations are presented in  \ref{AppA-ELderiv-sec}, 
which results in the relationship between probabilities $w_q(j)$ 
and fields $u_q(j)$ (for $q\in\{+1,0,-1\}$ and $1\leq j \leq N$) as 
\begin{align} 
u_+(j) - u_0(j) = & \;  \log w_+(j) - \log w_0(j) - \beta_+ + \beta_0 , \nn \\ 
u_-(j) - u_0(j) = & \;  \log w_-(j)  - \log w_0(j) - \beta_- + \beta_0 , 
\label{eqmot4}  \end{align}
together with the constraints (\ref{Con2})
\beq
N_+ = \sum_{j=1}^N w_+(j) , \quad N_- = \sum_{j=1}^N w_-(j) . 
\label{C2a}  \eeq

Rearranging (\ref{eqmot4}), we have 
$w_+ = w_0 \ee^{u_+ - u_0 + \beta_+ - \beta_0}$, 
$w_- = w_0 \ee^{u_- - u_0 + \beta_- - \beta_0 }$, and adding these to $w_0$, we find 
\begin{align}
w_+(j) =&\; \frac{\exp( \beta_+ \!+\! u_+(j))}{D } , \quad 
w_0(j) = \frac{\exp( \beta_0 \!+\!  u_0(j))}{D } , \quad 
w_-(j) = \frac{\exp(  \beta_- \!+\! u_-(j))}{D }  , \nn\\ 
D = & \; \exp( \beta_+ \!+\! u_+(j)) + \exp( \beta_0 \!+\! u_0(j))+ \exp( \beta_- \!+\! u_-(j)) 
. \label{w-gsol3}  \end{align}
% {\bf explain - Gibbs distn, lots more from ref comments} \\ % 4.10 
We observe that only the differences $u_+ - u_0$ and $u_- - u_0$ 
are relevant, and so just two fields will suffice.  In general, one can 
assume $u_0(j)=0$ for all $j$.   
Alternatively, in cases where only two genotypes ($\pm$) are present, we have 
\begin{align} 
w_+(j) = & \; \exp \left( \frac{ \beta_+ \!+\! u_+(j) }{2} \right) \sech \left( 
\frac{ u_+(j) -u_-(j) + \beta_+ - \beta _- }{2} \right) , \nn \\ 
w_-(j) = & \; \exp \left (\frac{ \beta_- \!+\! u_-(j) }{2} \right) \sech \left( 
\frac{ u_+(j) -u_-(j) + \beta_+ - \beta_- }{2} \right) . 
\end{align}
We will make use of these formulae later to determine the forms of the 
field strengths $u_q(j)$ and location probabilities $w_q(j)$ and their 
interdependencies.   \qb{By analogy with ergodic systems in statistical 
physics,  these expressions for the genotype location probabilities} 
(\ref{w-gsol3}) \qb{can be viewed as Gibbs distributions, where 
$\beta_q + u_q(j)$ takes the place of chemical potential}.

%{\bf explain $\tilde u$ } - removed here  % 16.9
Mathematically, we describe the {\em  forward} problem to be the 
determination of the observables, that is the path $\theta_\pm(j)$ 
and the cumulative distributions $W_q(j)$ from a given field $u_\pm(i)$.  
The {\em inverse} problem is defined to be the derivation of the field 
$u_\pm(i)$ from observed data for the path 
$\theta_\pm(j)$ and the distributions $W_q(j)$. 
Both formulations of the problem are complicated by the presence 
of Lagrange multipliers $\beta_\pm$. 
The inverse problem is simpler to solve, since (\ref{eqmot4}) 
can be rearranged to give independent and explicit expressions 
for the fields $u_\pm(j)$. 
If one considers the formulae (\ref{eqmot4}) as the forward problem 
for $w_\pm(j)$ the solution is complicated due to the coupling 
and the nonlinearity.  
In Section \ref{inv-sec} we illustrate the solution of a couple of 
cases of the inverse problem, finding the fields $u_\pm$ from given 
distributions $W_\pm$.  
In Section \ref{weak-sec}, we consider the forward problem 
in the case where the field-strengths $u_\pm$ are weak, 
that is small amplitude, but are nonzero and dependent on 
position $j$. 

%---------------------
% \subsection{{\bf Statistics of the random configuration} \label{rand-sec}}
%{\bf replace subsec by single sentence} % 4.1

If we consider the random configuration, in which there is no field imposed 
($u_q(j)=0$ for all $j$ and all $q$), then maximising the entropy gives the 
uniform distributions  $w_q(j) = w_q^{(0)}$. 
In this case, the expected values for the cumulative distributions (\ref{WWq-def}) 
are (\ref{WW0-def}). Thus, for a negative control gene or SNP (i.e.~one that 
has no causative effect or correlation with phenotype), the expected value of 
the $\theta$-paths are zero, that is, $\mathbb{E}[\theta_\pm(j)] = 0$ for all 
positions in the list $1 \leq j \leq N$. 

%Using the analytical approach with fields $u_\pm(j)=0$ implies $E=0$ and 
%from (\ref{w-gsol3}), we have $w_q(j)=\ee^{\beta_q}/(1+\ee^{\beta_+}+\ee^{\beta_-})$,
%which satisfies constraint (\ref{Con1}).   The second constraint (\ref{Con2}) 
%implies that in the average random configuration, $\beta_+,\beta_-$ are given by 
%$\ee^{\beta_q} = N_q/N_0$. 

%----------------------------------------------------------------------------
\section{{The inverse problem} \label{inv-sec}}
\setcounter{equation}{0}

Here we assume that the location probabilities $w_q(j)$, 
and hence the cumulative distribution $W_q(j)$ as well, are known 
functions, with $0 < w_q(j) < 1$ for $q \in \{ +1,0,-1\}$.   
We aim to determine the corresponding field-strengths $u_q(j)$,  
using (\ref{eqmot4}).  For the theory proposed in Section \ref{math-sec}, 
this inverse problem is more easily solved than the forward problem, 
whose analysis we delay to section \ref{weak-sec}. 

\qb{In this section, we consider a sample population taken from specific 
distributions $p(\Omega)$,  Thus we can consider our variables to be 
functions of phenotype value ($\Omega$) rather than position in the list ($j$);  
in place of} (\ref{WWq-def}),   \qb{the cumulative distribution of individuals 
of each genostate is then defined by } 
\begin{align} 
\tilde W_+(\Omega) = & \mbox{number of $+1$-genostate individuals 
	with phenotype $\leq \Omega $in the list $\Gamma$} , \nn\\ 
\tilde W_0(\Omega) = & \mbox{number of $0$-genostate individuals 
	with phenotype $\leq \Omega$ in the list $\Gamma$} , \nn\\ 
\tilde W_-(\Omega) = & \mbox{number of $-1$-genostate individuals 
	with phenotype $\leq \Omega$ in the list $\Gamma$} , 
\label{WWtq-def} 
\end{align} 
\qb{and we have the corresponding versions of all the variables, given by  
$\tilde W_q(\Omega(j)) = W_q(j)$,  
$\tilde \theta_q(\Omega(j)) = \theta_q(j)$,  $\tilde w_q(\Omega(j)) = w_q(j)$, 
$\tilde W_q^{(0)}(\Omega(j)) = W_q^{(0)}(j)$,  $\tilde w_q^{(0)} = w_q^{(0)}$, 
and $\tilde u_q(\Omega(j)) = u_q(j)$.  See} \ref{pheno-dep-sec} \qb{for more details.}

%----------------------------------------------------------------------------
\subsection{{\bf Gaussian (Normal) distributions}  \label{invG-sec1}}

Phenotypes are often  assumed to be normally distributed, that is, have 
a Gaussian distribution.   We assume that the distributions of $\{+1,0,-1\}$ 
states are given by the probability denisty functions 
\begin{align}
p_+(\Omega) = & \; N(\mu_+,\sigma_+), \qquad 
p_0(\Omega) = N(\mu_0,\sigma_0), \qquad 
p_-(\Omega) = N(\mu_-,\sigma_-) , \nn \\ 
N(\mu,\sigma) = & \; \frac{1}{\sqrt{2\pi}\sigma} \exp \left( 
- \frac{(\Omega-\mu)^2}{2\sigma^2} \right) , 
\end{align} 
where $\mu_q$ are the means of the distributions, generally taken to 
be distinct, and $\sigma_q$ the corresponding standard deviations, 
which could be distinct or the same.  Fisher \cite{Fisher,moran} 
typically assume them to have the same standard deviations.  

Assuming the sample has $N_+,N_0,N_-$ individuals of each 
corresponding genetic type, the probabilities $w_q(\Omega)$ 
of each position in the list being occupied by an individual of 
genotype $q \in \{+1,0,-1\}$ are given by 
\begin{align}
\tilde w_+(\Omega) = & \; \frac{ N_+ p_+(\Omega) }
{N_+ p_+(\Omega) + N_0 p_0(\Omega) + N_- p_-(\Omega)}, \quad 
\tilde w_0(\Omega) = \frac{ N_0 p_0(\Omega) }
{N_+ p_+(\Omega) + N_0 p_0(\Omega) + N_- p_-(\Omega)}, \nn \\
\tilde w_-(\Omega) = & \; \frac{ N_- p_-(\Omega) }
{N_+ p_+(\Omega) + N_0 p_0(\Omega) + N_- p_-(\Omega)} . 
\label{wtheory1} \end{align}

If we assume that $\sigma_+ = \sigma_- = \sigma_0 = \sigma$, 
then these formulae can be simplifed, to 
\begin{align}
\tilde w_+(\Omega) = &  \; \frac{ N_+ \exp( (\mu_0\!-\!\mu_+)
( \mu_0 \!+\!\mu_+ \!-\! 2\Omega ) / 2\sigma^2 ) }{ N_0 + 
N_+ \exp( (\mu_0\!-\!\mu_+) (\mu_0\!+\!\mu_+ \!-\!2\Omega ) / 2\sigma^2 ) + 
N_- \exp( (\mu_0\!-\!\mu_-)(\mu_0 \!+\! \mu_- \!-\!2\Omega) / 2\sigma^2 )  } , \nn \\ 
\tilde w_-(\Omega) = & \; \frac{ N_- \exp( (\mu_0\!-\!\mu_-)
( \mu_0 \!+\!\mu_- \!-\! 2\Omega ) / 2\sigma^2 ) }{ N_0 +  
N_+ \exp( (\mu_0\!-\!\mu_+) (\mu_0\!+\!\mu_+ \!-\!2\Omega ) 2/ \sigma^2 ) + 
N_- \exp( (\mu_0\!-\!\mu_-)(\mu_0 \!+\! \mu_- \!-\!2\Omega) / 2\sigma^2 )  } , \nn \\ 
\tilde w_0(\Omega) = & \;  \frac{ N_0 }{ N_0 +  
N_+ \exp( (\mu_0\!-\!\mu_+) (\mu_0\!+\!\mu_+ \!-\!2\Omega ) / 2\sigma^2 ) + 
N_- \exp( (\mu_0\!-\!\mu_-)(\mu_0 \!+\! \mu_- \!-\!2\Omega) / 2\sigma^2 )  } . 
\nn \\ & \label{Gaussp-wfn}
\end{align}

By comparing the above with (\ref{w-gsol3}), we see that the 
field strengths $\tilde u_\pm(\Omega)$ are given by 
\begin{align}
\tilde u_+(\Omega)  = & \; \log \left( \frac{N_+}{N_0} \right) + 
\frac{ (\mu_0-\mu_+)(\mu_0+\mu_+ - 2 \Omega) }{ 2\sigma^2} - \beta_+, \nn \\ 
\tilde u_-(\Omega)  = & \; \log \left( \frac{N_-}{N_0} \right) + 
\frac{ (\mu_0-\mu_-)(\mu_0+\mu_- - 2 \Omega) }{ 2\sigma^2 } - \beta_- , 
\label{uu-Gauss1}
\end{align}
where $\beta_\pm$ are constants (Lagrange multipliers). 
This calculation shows that if the phenotype distributions for the different 
genotypes are all normally distributed (Gaussians),  with different means, 
$\mu_q$, but share a common standard deviation (as assumed by Fisher 
\cite{Fisher,moran}) then the genotype field is {\em linear} in phenotype ($\Omega$), 
with $\tilde u_\pm = m \Omega + c$.  The gradient of the line ($m$) 
depends on the difference in means ($\mu_+-\mu_0$ and $\mu_--\mu_0$). 
Thus values of difference in means is influenced by the whole data set. 

This analysis has been for the case of general phenotype measurements, 
$\Omega$; the result for the case where we consider $w_\pm$ and $u_\pm$ 
be functions of position in the list, $j$, (rather than absolute phentotype value, 
$\Omega$) can be obtained simply by defining $\Omega(j) = j$. 

Whilst it is noteworthy that Gaussian distributions give rise to a linear field, 
there may be other phenotype distributions which also lead to linear fields, 
so the converse statement (that linear fields indicate normal distributions) 
is not necessariy true.  In fact, below, we show that another phenotype 
distribution also leads to linear field (see Section \ref{gamma-dist-sec}). 

%-----------
\subsection{{\bf More general Gaussian distribution} 
\label{mgenGdist-sec}}

If do not make the assumption that all the distributions 
$p_+(\Omega)$, $p_0(\Omega)$, $p_-(\Omega)$ have the same variance, 
then we do not obtain such a simple field dependence on $\Omega$.  
Following the same procedure as in Section \ref{invG-sec1}, denoting 
the standard deviations of the phenotype distributions by $\sigma_+$, 
$\sigma_0$, $\sigma_-$, we find 
\begin{align}
\tilde u_+(\Omega) = & \; \log \left( \frac{N_+ \sigma_0}{\sigma_+ N_0} \right) 
+ \frac{(\sigma_+^2-\sigma_0^2)\Omega^2}{2\sigma_+^2\sigma_0^2} 
+ \frac{ (\mu_+\sigma_0^2-\mu_0\sigma_+^2) \Omega}{\sigma_+^2\sigma_0^2} 
+ \frac{(\mu_0^2\sigma_+^2-\mu_+^2\sigma_0^2)}{2\sigma_+^2\sigma_0^2}  
 - \beta_+ , \nn \\ 
\tilde u_-(\Omega) = & \; \log \left( \frac{N_- \sigma_0}{\sigma_- N_0} \right) 
+ \frac{(\sigma_-^2-\sigma_0^2)\Omega^2}{2\sigma_-^2\sigma_0^2} 
+ \frac{ (\mu_-\sigma_0^2-\mu_0\sigma_-^2) \Omega}{\sigma_-^2\sigma_0^2} 
+ \frac{(\mu_0^2\sigma_-^2-\mu_-^2\sigma_0^2)}{2\sigma_-^2\sigma_0^2} 
 - \beta_- , 
\label{uu-Gauss2}
\end{align}
thus we see that the field-strength is now {\em quadratic} in phenotype 
value - still a relatively simple form, though not as simple as linear. 

%----------------------------------------------------------------------------
\subsection{{\bf Gamma distribution} \label{gamma-dist-sec}}

If we assume that the phenotypes for each genotype 
are distributed according to Gamma distributions, that is, 
\beq
p_q(\Omega) = \Omega^{k-1} \ee^{-\lambda \Omega} \lambda^k / \Gamma(k) , 
\eeq
with $q\in\{+1,0,-1\}$ denoting the genotypes, and parameters given by 
$k_+,k_0,k_-,\lambda_+,\lambda_0,\lambda_-$.   The relationships between 
the mean and standard deviation, and these parameters are  given by 
\beq
\mu = \frac{k}{\lambda} , \quad \sigma = \frac{\sqrt{k}}{\lambda} , \qquad 
k = \frac{\mu^2}{\sigma^2} , \quad \lambda = \frac{\mu}{\sigma^2} . 
\eeq
The location probabilities $w_+,w_0,w_-$ are given by (\ref{wtheory1})
which can be written as 
\beq
\tilde w_+(\Omega) = \frac{
N_+ \Omega^{k_+ -1} \ee^{-\lambda_+ \Omega} \lambda_+^{k_+} / \Gamma(k_+) }{
N_+ \Omega^{k_+ -1} \ee^{-\lambda_+ \Omega} \lambda_+^{k_+} / \Gamma(k_+) +  
N_0 \Omega^{k_0 -1} \ee^{-\lambda_0 \Omega} \lambda_0^{k_0} / \Gamma(k_0) +
N_- \Omega^{k_- -1} \ee^{-\lambda_- \Omega} \lambda_-^{k_-} / \Gamma(k_-) } ,  
\label{Gamma-genw}  \eeq
with similar formulae for $\tilde w_0(\Omega),\tilde w_-(\Omega)$. 

There are two special cases where some simplification occurs: 
(i) where the three genotypes share the same $k$, but have different $\lambda$ 
values, and (ii) where they share the same $\lambda$ and have different 
$k$-values.  In both cases, the means and the standard devations both differ.  
We consider each in turn. 

If we assume that $k_+=k_0=k_-=k$ then the formula (\ref{Gamma-genw}) 
simplies to  
\beq
\tilde w_+(\Omega) = \frac{ 
N_+ \ee^{(\lambda_0-\lambda_+) \Omega} (\lambda_+/\lambda_0)^{k}  }{ N_0 + 
N_+ \ee^{(\lambda_0-\lambda_+) \Omega} (\lambda_+/\lambda_0)^{k}  +  
N_- \ee^{(\lambda_0-\lambda_-) \Omega} (\lambda_-/\lambda_0)^{k} } ,  
\eeq
with similar formulae for $\tilde w_0,\tilde w_-$. 
Then comparing this expression with (\ref{w-gsol3}) gives
\begin{align}
\tilde u_+(\Omega) = & \; \log \left(\frac{N_+}{N_0} \right) 
+ (\lambda_0-\lambda_+) \Omega + 
k \log \left( \frac{\lambda_+}{\lambda_0} \right) - \beta_+ , \nn \\ 
\tilde u_-(\Omega) = & \; \log \left( \frac{N_-}{N_0} \right) 
+ (\lambda_0-\lambda_-) \Omega + 
k \log \left( \frac{\lambda_-}{\lambda_0} \right) - \beta_- , 
\end{align}
thus this case also corresponds to field strength ($\tilde u_\pm(\Omega)$) 
which varies {\em linearly} with phenotype value, $\Omega$. 
Since these distributions have different values of $\lambda$, both mean 
and standard deviation differ between the various genotypes. 

If we assume that the $k$'s are distinct, whilst  $\lambda_+ = \lambda_0 
= \lambda_- = \lambda$ then (\ref{Gamma-genw}) simplies to  
\beq
\tilde w_+(\Omega) = \frac{ 
(N_+/N_0) \Omega^{k_+ - k_0} \lambda^{k_+ - k_0} \Gamma(k_0) / \Gamma(k_+) }{1+
(N_+/N_0) \Omega^{k_+ - k_0} \lambda^{k_+ - k_0} \Gamma(k_0) / \Gamma(k_+) + 
(N_-/N_0) \Omega^{k_- - k_0} \lambda^{k_- - k_0} \Gamma(k_0) / \Gamma(k_-) } , 
\eeq
with similar formulae for $\tilde w_0,\tilde w_-$. 
Then comparing this expression with (\ref{w-gsol3}) gives
\begin{align}
\tilde u_+(\Omega) = & \; \log \left( \frac{N_+}{N_0} \right) 
+ (k_+ - k_0) \log (\Omega) + (k_+ - k_0)\log(\lambda) + 
\log \left( \frac{ \Gamma(k_0)}{\Gamma(k_+)} \right) - \beta_+ , \nn \\ 
\tilde u_-(\Omega) = & \; \log \left( \frac{N_-}{N_0} \right) 
+ (k_- - k_0) \log (\Omega) + (k_- - k_0) \log(\lambda) + 
\log \left( \frac{\Gamma(k_0)}{\Gamma(k_-)} \right) - \beta_- ,  
\end{align}
which corresponds to the field-strength ($\tilde u_\pm$) 
being logarithmic in phenotype value ($\Omega$). 

%-----------
\subsection{{\bf Form of $\theta-$path}}
\label{th-pred-sec}

From knowledge of the probabilities $w_q(j)$ \qb{or $\tilde w_q(\Omega)$},  
it possible to give formulae for the expected form of the $\theta$-paths, 
\qb{$\theta_q(j)$ or $\tilde \theta_q(\Omega)$}.  
We return to the case of Gaussian distributions with the same 
standard deviation studied in Section \ref{invG-sec1}; 
we assume that the means are close with respect to the standard 
deviation of the overall distribution ($\mu_\pm - \mu \ll \sigma$). 
This asymptotic relationship can be thought of either as the means of the 
three phenotype distributions being similar or the variance of all of them 
are large, so that the distributions strongly overlap. We write the means as 
\begin{align}
\mu_+ = \ol{\mu} + h \sigma \hat \mu_+ , \quad 
\mu_0 = \ol{\mu} + h \sigma \hat \mu_0 , \quad 
\mu_- = \ol{\mu} + h \sigma \hat \mu_- , \quad 
\mbox{with} \quad h  \ll1 , 
\end{align}
where the overall mean $\ol{\mu}$ is weighted by 
the number of each genotype in the sample 
\beq
\ol{\mu} = \frac{N_+ \mu_+ + N_0 \mu_0 + N_- \mu_-}
{N_+ + N_- + N_-} . 
\label{mubar}  \eeq
This condition (\ref{mubar}) implies that the perturbations 
$\mu_+,\mu_0,\mu_-$ satisfy 
\begin{align}
0 = N_+ \hat \mu_+ +  N_0 \hat \mu_0 + N_- \hat \mu_- .  
\label{muh-cond}
\end{align}

Expanding (\ref{wtheory1}) we obtain 
\begin{align}
\tilde w_+(\Omega) = & \; w_+^{(0)} +  h w_+^{(0)} \hat\mu_+ 
	(\Omega-\ol{\mu}) / \sigma , \nn \\ 
\tilde w_-(\Omega) = & \; w_-^{(0)} +  h w_-^{(0)} \hat\mu_- 
	(\Omega-\ol{\mu}) / \sigma , \nn \\ 
\tilde w_0(\Omega) = & \; w_0^{(0)} - h ( w_+^{(0)} \hat\mu_+ + 
	w_-^{(0)}\hat\mu_- ) (\Omega-\ol{\mu}) / \sigma , 
\label{whot}
\end{align}
which automatically satisfy the constraints
\beq
\tilde w_+(\Omega) + \tilde w_0(\Omega) + \tilde w_-(\Omega) = 1 , \qquad 
\int p(\Omega) \tilde w_q(\Omega) \, \dd \Omega = N_q , 
\eeq
for all $\Omega$ and any $q\in\{+1,0,-1\}$.  These conditions are met 
both at leading order (where $\tilde{w}_q(\Omega)=w_q^{(0)}$) and at 
$\mathcal{O}(h)$. The former constraing corresponds to $\sum_q w_q=1$ 
and the latter to $N_q = \sum_j w_q(j) = W_q(N)$. 
At $\mathcal{O}(h)$ we recover (\ref{muh-cond}) and 
$\int p(\Omega) (\Omega -\ol{\mu})\dd\Omega=0$, 
which is simply the definition of the mean of the distribution.  

Since $\tilde \theta_q(\Omega) = \tilde W_q(\Omega) - \tilde W_q^{(0)}(\Omega)$, 
by differentiating, and using  (\ref{C7})--(\ref{C9}) and (\ref{whot}), we obtain 
\beq
\frac{\dd \theta_q}{\dd j} 
= \frac{\dd \tilde \theta_q}{\dd \Omega} \frac{1}{Np(\Omega)} 
= \left( \frac{\dd \tilde W_q}{\dd \Omega} - \frac{\dd \tilde W_q^{(0)}}{\dd \Omega} 
	\right) \frac{\dd \Omega}{\dd j} 
=  w_q - w_q^{(0)} = h w_q^{(0)} \hat\mu_q(\Omega-\ol{\mu})/\sigma , 
\label{thode}
\eeq
Since $p(\Omega)$ is Gaussian, it satisfies the ordinary differential 
equation $p' = - (\Omega-\ol{\mu})p /\sigma^2$, hence we rearrange (\ref{thode}) 
and solve 
\beq
\frac{\dd \tilde \theta_q}{\dd \Omega} = h w_q^{(0)} \hat \mu_q N p(\Omega) 
\left( \frac{\Omega-\ol{\mu}}{\sigma} \right) 
= - h w_q^{(0)} \hat \mu_+ \sigma N \frac{\dd p}{\dd \Omega} 
\eeq
by $\tilde \theta_q (\Omega) = h \sigma N w_q^{(0)} \hat \mu_q p(\Omega)$, 
which clearly has the properties 
$\tilde\theta\rightarrow0$ as $\Omega \rightarrow \pm \infty$, 
and $\dd\tilde\theta/\dd\Omega=0$ at the mean of $p(\Omega)$. 
Also, we expect the general magnitude of $\tilde\theta$ to be proportional 
to the the difference in means ($\hat\mu_q$),  
and the number of genotype $N_q = N w_q^{(0)}$ as in 
\beq
\tilde \theta_+(\Omega) = N_+ (\mu_+ - \ol{\mu}) p(\Omega) , \quad 
\tilde \theta_-(\Omega) = N_- (\mu_- - \ol{\mu}) p(\Omega) , \quad 
\tilde \theta_0(\Omega) = N_0 (\mu_0 - \ol{\mu}) p(\Omega) . 
\eeq
For other distributions, that is, $p$ not Gaussian with identical variances, 
there is no reason for $\tilde\theta$ to follow the $p$, since 
$\tilde\theta'(\Omega) \propto (\Omega-\ol\mu) p \neq p'(\Omega)$. 
For example, if the distributions $p_q(\Omega)$ are Gaussian with 
identical means but different variances, the $\theta$-paths 
may be positive in some ranges of $\Omega$ and negative 
elsewhere (as illustrated in the next subsection). 

%---------------------------
\subsection{{\bf Effect of different standard deviations}} 

We now consider the case of the phenotype distributions of 
the various genotypes having the same mean but slightly different 
variances. Thus we have 
\begin{align} 
p(\Omega) = \frac{1}{\sigma \sqrt{2\pi}} \exp \left( - 
\frac{ (\Omega-\mu)^2 }{2\sigma^2} \right) , 
\end{align}
with $\mu$ the same for all genotypes $q\in \{+1,0,-1\}$ and standard 
deviations given by $s_q = \ol{s} + h \hat{s}_q$, where $h \ll 1$, and 
$\ol{s}$ is chosen by $\ol{s} = (1/N) \sum_q N_q s_q$ so that 
$\sum_q N_q \hat{s}_q=0$.  
The phenotype distribution for the three genotypes and the location 
probabilities $\tilde{w}_q = N_q p_q/ \sum_{q'} N_{q'} p_{q'}$ are given by 
\begin{align} 
p_q(\Omega) = p(\Omega) \left[ 1 + \frac{ h \hat{s}_q }{ \ol{s}^2 } 
\left( (\Omega -\mu)^2 - \ol{s}^2 \right)  \right] , 
\qquad 
\tilde{w}_q(\Omega) = w_q^{(0)} + w_q^{(0)} \frac{ h \hat{s}_q}{\ol{s}^2} 
\left[ (\Omega-\mu)^2 - \ol{s}^2 \right] . 
\end{align}
Using (\ref{thode}), we have 
\beq
\frac{\dd \tilde \theta_q}{\dd \Omega}  = w_q - w_q^{(0)} = 
N p(\Omega) w_q^{(0)} h \hat{s}_q \ol{s}^{-2} [ (\Omega-\mu)^2  - \ol{s}^2 ] , 
\eeq
which is solved by 
\beq
\tilde\theta_q(\Omega) = - \frac{N w_q^{(0)} h \hat{s}_q (\Omega-\mu)}
{ \ol{s} \sqrt{2\pi}} \exp\left( - \frac{(\Omega-\mu)^2}{2\ol{s}^2} \right) . 
\eeq
This function changes sign at $\Omega=\mu$, whilst approaching zero 
in both the limits $\Omega \rightarrow\pm\infty$.

%---------------------------
\subsection{{\bf Summary}}

\qb{In Sections} \ref{math-sec} and \ref{inv-sec} \qb{ we have generalised 
Shannon's information theory to include phenotypic fields, $u_q(j)$ which 
account for and describes the effects that genotype has on the ordered 
phenotype list. 
For example, if the effect of the genotype is that the $+1$-genostate 
have values of $\Omega$ larger than the $0$-genostate, and the 
$-1$-genstate are smaller, then we see that the field $\tilde u_+(\Omega)$ 
is an increasing function of $\Omega$ and $\tilde u_-(\Omega)$ is a 
decreasing function. More specifically 
}    
for general phenotypic distributions for each of the genotypes of the form 
$p_+(\Omega)$, $p_0(\Omega)$, $p_-(\Omega)$ we have 
the field strength being given by 
\begin{align}
\tilde u_+(\Omega) =  \log \left( \frac{N_+}{N_0} \right)  + \log \left( 
\frac{p_+(\Omega)}{p_0(\Omega)} \right) - \beta_+ , & \quad 
\tilde u_-(\Omega) =   \log \left( \frac{N_-}{N_0} \right)  + \log \left( 
\frac{p_-(\Omega)}{p_0(\Omega)} \right) - \beta_- . 
\end{align}
We have considered various forms for the location probabilities 
$\tilde w_q(\Omega)$, and in each case found explicit formulae for the 
field strengths, $\tilde u_\pm(\Omega(j))$, highlighting some special cases 
where the fields are linear in the phenotypes, due to differences 
in the distributions of the phenotype for each genotype.

%----------------------------------------------------------------------------
\section{{The forward problem} \label{weak-sec}}
\setcounter{equation}{0}

Having examined a few examples of the calculation deriving 
field strengths $u_q(j)$, from the location probabilities $w_q(j)$,  
we return to the mathematically more difficult problem of determining 
the location probabilities $w_q(j)$ or $\tilde w_q(\Omega(j))$ from the 
field strengths $u_\pm(j)$ or $\tilde u_\pm(\Omega(j))$. 
In general this is a nonlinear problem, due to the gobal constraints 
on the total number of each genotype in the distributions. 
Hence we focus on generating an approximate solution, in the case 
of a weak but nonzero dependence of phenotype on genotype, 
this corresponds to the field terms $u_q(j)$ being small. 

The algebra is simplied by introducing constants $A=\ee^{\beta_+}$ 
and $B = \ee^{\beta_-}$, and an interaction `potential' $v_\pm(j)$ 
given by $v_\pm(j) = \exp u_{\pm}(j)$ so that we have algebraic 
relationships between $v_\pm$ and $w_\pm$.  In this notation, 
the zero genetic state $u_0(j)=0$ corresponds to $v_0(j) = \ee^u_0=1$. 
From (\ref{w-gsol3}) we have 
\begin{align}
w_0(j) =&\; \frac{1}{ 1 + \ee^{ \beta_+ + u_+(j) } + \ee^{\beta_- + u_-(j)}  }  
= \frac{1}{1+Av_+(j) + B v_-(j)} , \nn \\ 
w_+(j) =& \; \frac{\ee^{ \beta_+ + u_+(j)} }
{1 + \ee^{ \beta_+ + u_+(j) } + \ee^{\beta_- + u_-(j)} } 
= \frac{A v_+(j)}{1 + A v_+(j) + B v_-(j)}  , \nn \\ 
w_-(j) = & \; \frac{\ee^{\beta_- + u_-(j) } }
{1 + \ee^{ \beta_+ + u_+(j) } + \ee^{\beta_- + u_-(j)} } 
= \frac{B v_-(j)}{1 + A v_+(j) + B v_-(j)}  ,  \label{w-gsol4}  
\end{align}
with $A,B$ determined by (\ref{C2a}), namely 
\begin{align}
N_+ = & \; \sum_{j=1}^N w_+(j) = \sum_{j=1}^N 
\frac{ A v_+(j) }{ 1 + Av_+(j) + B v_-(j) } , \nn \\
N_- = & \;  \sum_{j=1}^N w_-(j) = \sum_{j=1}^N 
\frac{ B v_-(j) }{ 1 + A v_+(j) + B  v_-(j) } .   
\label{C2b}\end{align}
Whilst the $j$-dependence is given relatively straightforwardly 
by (\ref{w-gsol4}), the problem of determining $A,B$ from the 
nonlinear equations (\ref{C2b}) is the main complicating factor. 

%------------------------------------------------------
\subsection{{\bf Weak-field analysis}}

Since we are aiming to solve the system (\ref{C2b}) with (\ref{w-gsol4}) 
in the weak field limit, we introduce a small parameter, $\ep \ll1$ and 
assume $u_\pm \sim \ep$, so that $v_\pm(j) = 1 + u_\pm(j) + \mathcal{O}(\ep^2)$.   
We expect the probabilities $w_\pm(j)$ to be close to $w_\pm^{(0)}$, 
which, from (\ref{w-gsol3}) with $u_q=0$, are given by 
\beq
w_+^{(0)} = \frac{A}{1+A+B} , \quad w_-^{(0)} = \frac{B}{1+A+B} , 
\quad w_0^{(0)} = \frac{1}{1+A+B} , 
\label{lot-wo-sol}
\eeq
and are solved by  
\beq
A = \frac{ w_+^{(0)} }{ 1 - w_+^{(0)} - w_-^{(0)} } , \qquad 
B = \frac{ w_-^{(0)} }{ 1 - w_+^{(0)} - w_-^{(0)} } . 
\label{ABlot} \eeq 
Note that $A,B$ are $\mathcal{O}(1)$ quantities.  

Expanding (\ref{w-gsol4}) to $\mathcal{O}(\ep)$, we find 
\begin{align}
w_+(i) = & \; \frac{A}{1\!+\!A\!+\!B} \left[ 1 + 
\frac{(1\!+\!B)u_+(i)}{1\!+\!A\!+\!B} - \frac{B u_-(i)}{1\!+\!A\!+\!B} \right] 
+ \mathcal{O}(\ep^2) ,  \nn \\ 
w_-(i) = & \; \frac{A}{1\!+\!A\!+\!B} \left[ 1 + 
\frac{(1\!+\!A)u_-(i)}{1\!+\!A\!+\!B} - \frac{A u_+(i)}{1\!+\!A\!+\!B} \right] 
+ \mathcal{O}(\ep^2) . 
\end{align}
with the constraints 
\begin{align}
N_+  = & \;  \sum_{j=1}^N w_+(j) =  \frac{ N A  }{ 1\!+\!A\!+\!B  } + 
\frac{A}{1\!+\!A\!+\!B} \sum_{j=1}^N \left( 
\frac{(1\!+\!B) u_+(j) - B u_-(j) }{1+A+B} \right) + O(\ep^2) , \nn \\  
N_-  = &\;  \sum_{j=1}^N w_-(j) =  \frac{ N B   }{ 1\!+\!A\!+\!B  } + 
\frac{B}{1\!+\!A\!+\!B} \sum_{j=1}^N \left( 
 \frac{- A u_+(j) + (1\!+\!A) u_-(j) }{1+A+B} \right) + O(\ep^2) .
\label{asy27}  \end{align} 
We now aim to find solutions for $A,B$ in terms of a series write 
\beq
A = A_0 + \ep A_1 + \ldots , \qquad B = B_0 + \ep B_1 + \ldots  
\eeq
with $A_k,B_k = O(1)$. 

After substiting these expansions into (\ref{asy27}), at leading order, 
we find $A_0,B_0$ are given by $A_0=N_+/N_0$ and $B_0=N_-/N_0$ 
as in (\ref{ABlot}).   This solution describes the uniform distribution 
of the genotypes across the range of phenotypes as in the random case. 
To gain insight into the effect of the field, we consider the next order 
terms in this expansion. 

At $\mathcal{O}(\ep)$ we find (\ref{asy27}) are solved by 
\beq
\ep A_1 = - A_0 \ol{u}_+ , \quad \ep B_1 = - B_0 \ol{u}_- , \quad 
\ol{u}_+ = \frac{1}{N} \sum_{i=1}^N u_+(i) , \quad 
\ol{u}_- = \frac{1}{N}\sum_{i=1}^N u_-(i) , 
\eeq
where $\ol{u}_\pm$ are the average strength of the field $u_\pm(j)$ 
over all locations $j$.   The resulting probabilities are defined by 
\beqa
w_+(i) & = & \frac{A_0}{D} \left( 1 + u_+(i) - \ol{u}_+ \right) , \quad 
w_-(i) = \frac{B_0}{D} \left( 1 + u_-(i) - \ol{u}_- \right) , \nn \\ 
D & =& 1 + A_0 + B_0 + A_0(u_+(j)-\ol{u}_+) + B_0 (u_-(i)-\ol{u}_-) 
\eeqa
which can be rewritten as 
\begin{align}
w_+(j)  = & \; w_+^{(0)} + w_+^{(0)}(1-w_+^{(0)})(u_+(j)-\ol{u}_+) 
	- w_+^{(0)} w_-^{(0)}(u_-(j)-\ol{u}_-)  , \nn\\ 
w_-(j) = & \;  w_-^{(0)} + w_-^{(0)}(1-w_-^{(0)})(u_-(j)-\ol{u}_-) 
	- w_-^{(0)}w_+^{(0)} (u_+(j)-\ol{u}_+ )  . \label{2t-wpm}
\end{align}
We note that both the fields $u_\pm(j)$ influence both the location 
probabilities $w_\pm(j)$; \qb{(hence, both also influence $w_0(j)$)}. 
% {\bf improve english } % 24.13
The important factors \qb{are the differences between} 
the local fields \qb{$u_\pm(j)$ and} the mean values \qb{$\ol{u}_\pm$}. 
These differences are further modulated by the relative numbers 
of $+1$ and $-1$ genetic states in the population $w_+^{(0)}=N_+/N$, 
$w_-^{(0)} = N_-/N$, with these coefficients dropping to zero if 
there are no states or if all entries have the same state.  
The local probabilities experience their largest values when there 
are similar numbers of the states present.   
This solves the problem of determining the location probabilities 
$w_q(j)$ and hence the expected cumulative distribution functions 
$W_q(j)$, from the field strengths $u_q(j)$.  

However, it is possible to take these calculations further, and give 
predictions for the form of the $\theta_\pm(j)$ functions.  
Since the cumulative distributions are given by 
\beq
W_+(j) = \sum_{i=1}^j w_+(i) , \quad 
W_0(j) = \sum_{i=1}^j w_0(i) , \quad 
W_-(j) = \sum_{i=1}^j w_-(i) , 
\eeq
and the $\theta_\pm$-paths by (\ref{Th-def}), we have 
\begin{align}
\theta_+(j) = & \; w_+^{(0)}(1-w_+^{(0)})\left( \sum_{i=1}^j u_+(i) - j \ol{u}_+ \right) 
	- w_+^{(0)} w_-^{(0)} \left( \sum_{i=1}^j u_-(i) - j \ol{u}_-  \right) , \nn \\ 
\theta_-(j) = & \; w_-^{(0)}(1-w_-^{(0)}) \left( \sum_{i=1}^j u_-(i) - j \ol{u}_- \right) 
	- w_-^{(0)}w_+^{(0)} \left( \sum_{i=1}^j u_+(i) - j \ol{u}_+ \right) . \label{2t-TH}
\end{align}
Note that all four terms in large brackets are zero at $j=0$ and $j=N$, 
but can be positive or negative for intermediate values ($1\leq j < N$).  
The leading order solution (\ref{2t-wpm}) is the same as the random case 
giving $\theta_\pm$-paths which are the same as random case,
that is,  $\theta_\pm(j)\equiv 0$ for all $i$.  The terms present in (\ref{2t-TH}) 
are all of $\mathcal{O}(\ep)$ in magnitude.

%----------------------------------------------------------------------------
\section{{Numerical results} \label{num-sec}}
\setcounter{equation}{0}

We illustrate the method using two sources of data, first we illustrate the 
method using synthetic data, taking samples from known distributions 
so that expected values can be quoted as well as calculated.  
Secondly, we use sample data from {\it arabidopsis thaliana}. 

%----------------------------------------------------------------------------
\subsection{{\bf Illustration using synthetic data}}

\begin{figure} % FIG 5 
\hspace*{-10mm}
\includegraphics[width=1.1\textwidth]{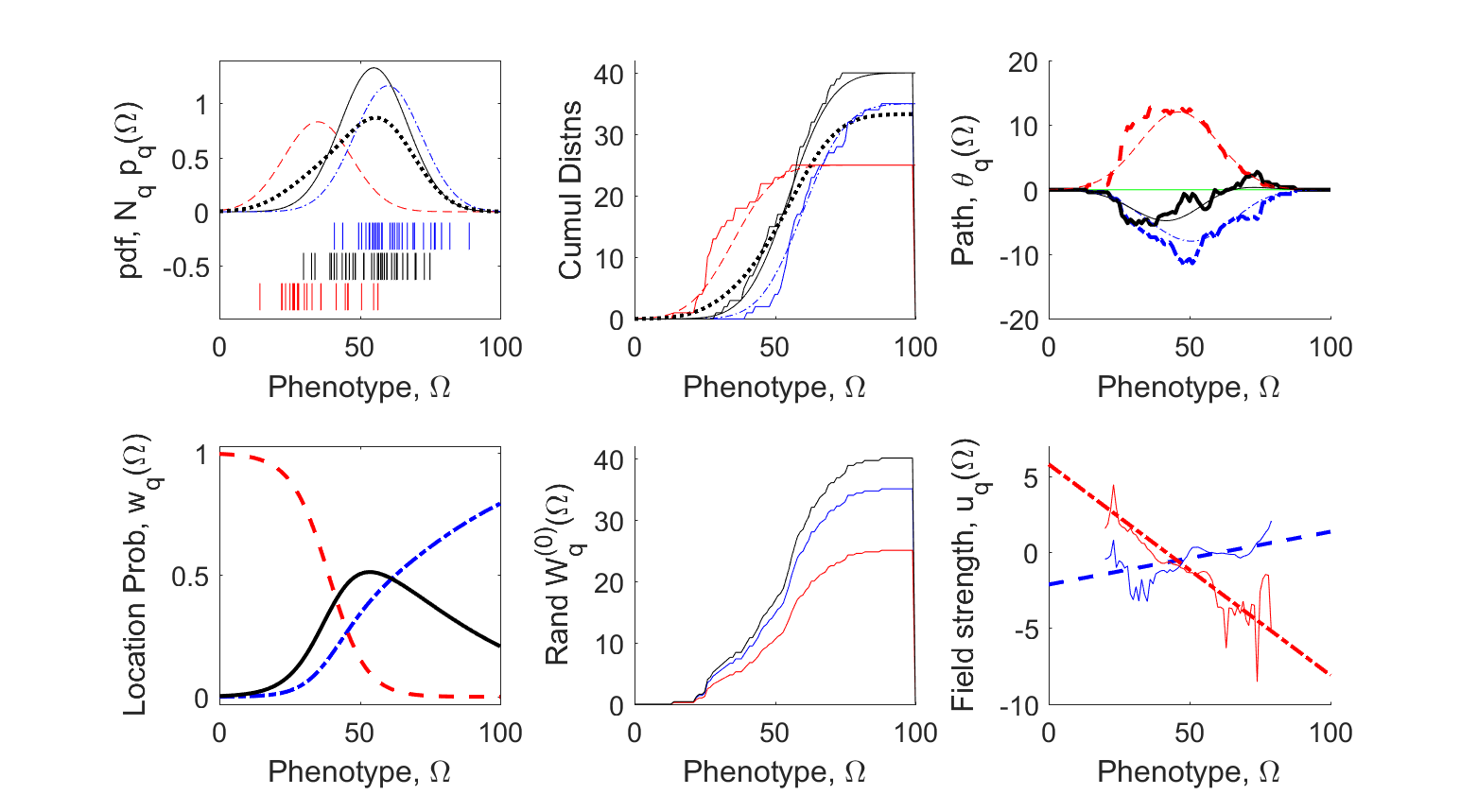}
\caption{ Top  left, the upper part shows that phenotype distributions 
of $+1,0,-1$ genetic states, 
`bar codes' at the bottom show the sample actually used. 
Dashed red curve corresponds to the $-1$ genetic state (bottom bar code); 
solid black curve corresponds to the $0$-state (middle bar code), 
dash-dotted blue curve corresponds to the $+1$  state (upper bar code), 
the dotted black line indictates the overall phenotype distribution, 
scaled down by a factor of three.  
Top centre: the cumulative distributions of each state ($\tilde W_q(\Omega(j))$), 
both expected values (smooth curves) and actual curves from the sample. 
Top right panel: plots of $\theta_q(j)$ obtained from the difference 
($\tilde W_q - \tilde W_q^{(0)}$); narrower lines show expected values. 
Lower left panel: location probabilities, $\tilde w_q(j)$ from (\ref{wtheory1}). 
Lower centre panel: the cumulative distribution assuming 
a random allocation of the samples. 
Lower right panel: field strengths $u_\pm(\Omega)$, 
expected (theoretical) values given by (\ref{eqmot4}), 
and calculated values given by (\ref{ucalc}). 
\label{num-fig} }
\end{figure}

Here, we assume that the phenotype distributions of each genotype 
($q\in\{+1,0,-1\}$) is given by a normal (Gaussian) distribution, 
with distinct means 
\beq
\Omega \sim  N( \mu_q, \sigma_q ) = p_q(\Omega) . 
\label{synth}
\eeq
Whilst the standard deviations could be distinct, 
the illustrative calculations given in Figure \ref{num-fig}
are for a case in which the standard deviations are all the same. 
In particular, the results are for the parameter values 
\begin{align}  
\mu_+  = & 60 , &  \mu_0 = &55 , & \mu_- =& 35,& \sigma = & 12 , 
\nn \\  N_+ = & 35 , & N_0 = & 40, \qquad& N_- = & 25 .&& 
\label{SampleParams}
\end{align}
The probability density distributions $p_q(\Omega)$ are illustrated in 
the top left panel of Figure \ref{num-fig}.
These values are chosen to test the method in several ways and illustrate 
a variety of outputs: 
whilst the +1 and 0 states have similar means, these 
differ substantially from the -1 state; the overall phenotype distribution 
if far from normal.  

Using the sample data, illustrated by the `bar-codes' in the lower part of 
panel 1 in Figure \ref{num-fig}, we construct cumulative distributions $\tilde 
W_q(\Omega)$,  which are illustrated in the top centre panel; specifically 
\beq
\tilde W_q(\Omega) = \mbox{number of individuals with $<\Omega$ 
and with phenotype $q$} . 
\eeq
In addition, we construct the cumulative distribution of the whole 
sample, $P_\Omega(\Omega) = ( N_+ P_+(\Omega) + N_0 P_0(\Omega) 
+ N_- P_-(\Omega))/N$. 
We then consider the random allocation of genetic states, and construct the 
$\theta$-paths $\theta_q(\Omega) = W_q(\Omega)-W_q^{(0)}(\Omega)$, 
which is illustrated in the top right panel.  
The cumulative distributions for the averaged random configuration 
are shown in the lower centre panel.  
In the top right panel, we also plot the expected value of the $\theta$-paths, 
obtained by evaluating  
$\theta_q(\Omega) = N_q P_q(\Omega) - N_q P_\Omega(\Omega)$. 

The predicted location probabilities $w_q(j) = \tilde w_q(\Omega(j))$ given by 
(\ref{wtheory1}) are plotted in the lower left panel.   These can be derived 
from $\theta_q(j) = \tilde \theta_q(\Omega(j))$ using $\tilde W_q = 
\tilde W_q^{(0)} + \theta_q$, equations (\ref{C7}), and (\ref{C9})  
\beq
w_q = \frac{\dd W_q}{\dd j} 
%= \frac{\dd \theta_q(j)}{\dd j} + \frac{\dd W_q^{(0)}}{\dd j} 
= w_q^{(0)} + \frac{1}{N p(\Omega)} \frac{\dd \theta_q}{\dd \Omega} , 
\label{w-theta}
\eeq
which can be used to produce the field strength by (\ref{eqmot4}) 
\begin{align}
\tilde u_+(\Omega) = &\;\ln \left(  \frac{
N p(\Omega) w_+^{(0)} + \dd \theta_+/\dd \Omega}{
N p(\Omega) w_0^{(0)} + \dd \theta_0/\dd \Omega} \right) , & \quad  
\tilde u_-(\Omega) = &\;\ln \left(  \frac{
N p(\Omega) w_-^{(0)} + \dd \theta_-/\dd \Omega}{
N p(\Omega) w_0^{(0)} + \dd \theta_0/\dd \Omega} \right) .  
\label{ucalc} \end{align}
As can be seen in the lower right panel of figure \ref{num-fig}, 
the numerical evaluation of a derivative leads to an increase 
in the noise.  However, the solid narrow curves show relatively 
good fits to a straight line in the range of phenotype where there 
is a larger amount of data - namely - for smaller phenotype 
values for $u_-$ (the red curve) and larger phenotype values 
for $u_+$ (the blue curve).   There are many approaches 
which could be used to smooth the $\theta$-data 
before taking the derivatives, for example, local averaging, 
or binning data. 
The blue dashed and red dash-dotted lines correspond to the 
theoretical values, these are linear due to the assumption 
of Gaussian distributions of phenotype values for the three 
genotypes, and these having the same standard deviation, 
that is $\sigma_q=\sigma$ for all of $q=+1,0,-1$ in (\ref{synth}).  
If more general distributions are used, the field strengths 
have more general shapes. 

%----------------------------------------------------------------------------
\subsection{{\bf Analysis of data from arabidopsis }}

Here we consider a subset of SNPs from arabidopsis thaliana 
\cite{sian-data} and apply the method outlined in sections 
\ref{deriv-sec} and \ref{pval-sec} to calculate the significance 
parameters (\ref{psnpmin})--(\ref{psnp5}). 

\begin{figure} % FIG 6
\includegraphics[scale=0.5]{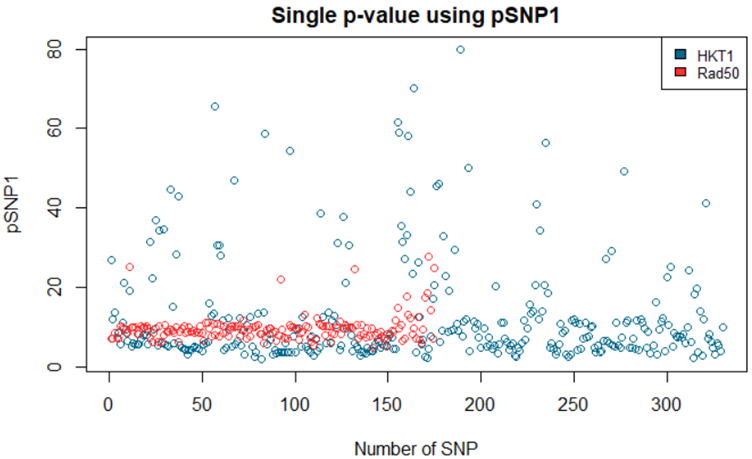}
\includegraphics[scale=0.5]{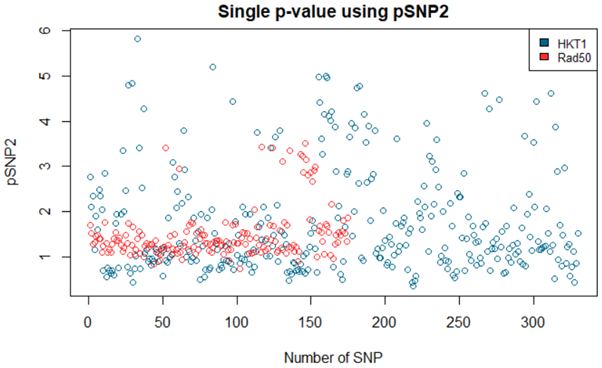}
\includegraphics[scale=0.5]{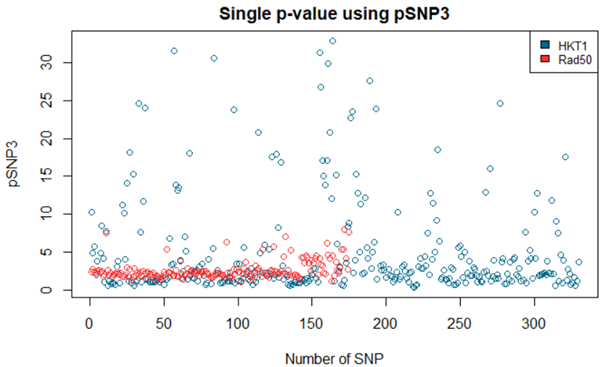}
\includegraphics[scale=0.5]{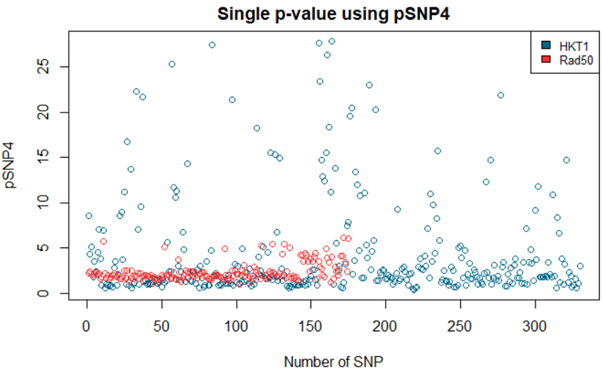}
\includegraphics[scale=0.5]{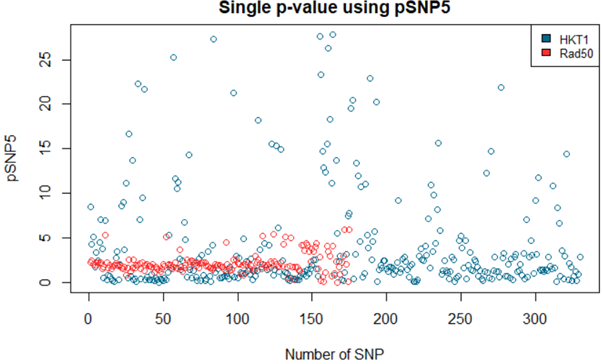}
\caption{Plots of $-\log p$ values for a range of SNPs over 
a gene that is known to be significant ($\sim330$ SNPS from HKT1, 
shown in blue online), and a negative control gene, ($\sim 170$ SNPs 
from Rad50, shown in red online). 
We illustrate 5 different methods for averaging the $p$ values from 
the list $1 \leq j \leq N$ to obtain a single $p$-value for the whole SNP. 
Top left: $-\log p$-value for SNP using the minimum (\ref{psnpmin}); 
top right: $-\log p$-value calculated using the mean $p$-value 
over the list (\ref{psnp2}); 
centre left: (\ref{psnp3}) the average of $-\log p$-values; 
centre right: using the mean of scaled $|\theta|$, (\ref{psnp4});
lower left: using the mean of scaled $\theta$ (\ref{psnp5}). 
\label{giota-fig}}
\end{figure}

We illustrate the output from two case studies: 
firstly, we consider about 330 SNPs from the HKT1 gene, 
which is known to be a significant in the uptake of sodium, 
so we would expect a strong correlation between 
sodium levels and certain SNPs on the HKT1 gene. 
The second case is the RAD50 gene, which is involved 
with detection of damage in DNA, and subsequent repair. 
There is no reason to expect this to be correlated to ion uptake; 
so this provides a negative control, only $\sim$170 SNPs 
on this gene are considered. For both of these genes, 
only two genetic states are present in the sample, 
corresponding to $+1$ and $-1$; there are no zero-states. 

In Figure \ref{giota-fig} we plot the $p$-values for each SNP, with each 
panel illustrating a different measure (\ref{psnpmin})--(\ref{psnp5}).  
Results for both genes are shown in the same panel, HKT1 in blue 
and RAD50 in red. 
Panel 1 shows that using the minimum value of $p(j)$ (\ref{psnpmin}) 
gives a reasonable separation between SNPs which do not 
influence phenotype and those that do, although about SNP 150-170 
we see several RAD50 SNPs which are moderately significant. 
For some SNPS, the $-\ln p$ values in this case are quite extreme. 
Panel 2 has much smaller $p$-values across the whole range, 
and shows more RAD50 SNPs as significant, hence we conclude 
that taking the mean $p$-value across all positions in the list as 
in (\ref{psnp2}) is a poor indicator of significance. 

The other three measures (\ref{psnp3})--(\ref{psnp5}) all show 
very similar and very good results.
Equation (\ref{psnp3}) corresponds to taking the average of the 
$ln p(j)$-values;  in (\ref{psnp5}) we take a weighted average of 
$\theta_q(j)$ values and use that to compute a $p$-value; 
(\ref{psnp5}) is similar to (\ref{psnp4}), but uses a weighted 
average of $|\theta|$ values.  In all these cases, 
the $-\ln (p)$-values range from zero to $\sim$30, 
and the significance of RAD50 SNPs all lie in the range $<10$, 
with a scatter of HKT1 SNPs being much more significant. 
These points are clustered around SNPs 25-30,  150-170, 240, 310. 

%----------------------------------------------------------------------------
\section{{Conclusions} \label{conc-sec}}
\setcounter{equation}{0}

We have outlined a \qb{Genomic Informational Field Theory (GIFT)}  
which combines knowledge of the genotypes of a population sample with 
a ranked list of phenotype values to extract information on the strength 
of interaction between genotype and phenotype. 
This can be applied to any continuous phenotype measurements, 
and used across a range of SNPs to determine those which 
have greatest influence on a particular characteristic.  
Such analyses will be the topic of future work \cite{bray}.  

We have derived formulae for the calculation of $p$-values in 
both the biallelic case (3 genetic states labelled $+1,0,-1$) 
and the mono allelic case (only $+$ and $-$ states). 
Both derivations a continuum limit of the theory which requires 
a large value of $N$ -- the number of individuals in the sample, 
and reasonably large number of each genetic state. 
\qb{ These $p$-values, together with a choice of significance 
level (e.g.~5\% or 1\%) and false discovery rate correction factors, 
enable one to determine which mutations have a significant impact 
on phenotype (measured physical characteristic)}. 
The model makes no assumptions on the form of the data - it 
may or may not exhibit Gaussian distribution, it may or may 
not fit the Hardy-Weinburg assumption ($N_0^2=4N_+N_-$). 
However, in Section \ref{inv-sec} we find a few special 
properties that hold in the case of phenotype distributions 
which are Gaussian, in particular, if the distributions for the 
various genotypes have the same variance, and similar means, 
then the field strength is approximately linear in phenotype value, 
and the shape $\theta-$path is simply a multiple of the rescaled 
Gaussian distribution.   For more general distributions these 
properties are no longer hold, but the shape of the $\theta$ 
trajectory and the form of the field are still meaningful. 

The mathematics underlying the model relies on a combination 
of Shannon's Information theory \cite{Shannon} and variational 
calculus \cite{gold} to relate information content to a postulated 
field which describes the relationship between genotype and 
phenotype. We outline how to determine field-strength from data. 
Preliminary numericaly studies show this method highlights more 
genes as having an influence on phenotype than classic GWAS; 
this is due to the ability to distinguish between negative controls 
(SNPs which have no influence on phenotype) and SNPS which 
have a weak influence.  \qb{ In the cases where genotype 
influences phenotype, we have introduced a field $u_\pm(j)$ 
which quantifies the strength and form of interaction between 
genetic states $\pm1,0$ and phenotype.  Preliminary results, 
both theoretical and numerical, show that monotone fields are 
due to genotype causing a shft in the mean of the phenotype 
distribution between different genotypes, as illustrated in}  
(\ref{uu-Gauss1}) and Figure \ref{num-fig}, \qb{whilst 
more general fields can indicate the genotype causing 
a change in standard deviation of the phenotype distributions} 
(\ref{uu-Gauss2}). %% . 

In future work, we propose to use these techniques to study the 
genome of arabidopsis \cite{bray}, includeing SNPs which are 
genuinely bi-allelic - that is - where the zero state is present in the 
sample, along with with the +1 and -1 states, and use larger 
samples, so that the informational fields can be explored. 
We also propose to study in more detail the relationship 
between these methods and the work of Fisher, to explore 
the various sources of variance in phenotype values, 
and to analyse cases where a single field can be used to analyse 
the phenotype-genotype correlation \cite{rauch}.

%----------------------------------------------------------------------------
\subsection*{{\bf Acknowledgements}}

\qb{We thank the referees for their helpful suggestions on ways to 
improve the manuscript}. 
We are grateful to the UKRI-funded Physics of Life network  for 
funding the work of PK. 

%----------------------------------------------------------------------------
\subsection*{{\bf Conflict of interest}}

The authors declare that they have no conflict of interest.

%----------------------------------------------------------------------------
\appendix
\renewcommand{\theequation}{\Alph{section}\arabic{equation}}
\setcounter{equation}{0}

\section{Calculation of the Variational Derivative 
\label{AppA-ELderiv-sec}}
\setcounter{equation}{0}

The equation relating the field strength $u_\pm$ to the genotype distribution 
functions $w_q(j)$ in Section \ref{math-sec} comes from a variational derivative. 
We solve this is Euler-Lagrange problem with constraints using Lagrange multipliers. 
We wish to find stationary points of the action (\ref{ActionA}) 
\begin{align}
\mathcal{A}[ \bw,\balpha,\bbeta] = & \; S[\bw] - E[\bw] + \sum_{j=1}^N \alpha_j C_j(\bw) 
+ \sum_{r\in\{+1,0,-1\} } \beta_r C_r(\bw) , 
\end{align} 
where the terms $C_*(\bw)$ represent constraints of the form $C_*(\bw)=0$ 
that must be satisfied, as detailed in equation (\ref{Con1}), and $\bw = 
(w_+(j),w_0(j),w_-(j))$, $\balpha = (\alpha_1,\alpha_2, \ldots,\alpha_N)$, 
$\bbeta = (\beta_+,\beta_0,\beta_-)$. 

To derive the corresponding constrained Euler-Lagrange equations, we allow 
all the local genotype probabilities $\mathbf{w}$ to be perturbed from $\bw = 
(w_+(j),w_0(j),w_-(j))$ to $\bw+h\bdelta$, 
where $h\ll1$ is a small scalar quantity, and $\bdelta = (\delta_+(j), \delta_0(j), 
\delta_-(j))$ are general $\mathcal{O}(1)$ perturbations, which satisfy certain 
contraints that will be derived later (\ref{de-con}).  

We consider the first Fr\'{e}chet derivative of $\mathcal{A}[\bw,\balpha]$, 
which is the difference in the Action between the perturbed state and the original 
state in the limit of small $h$, that is 
\begin{align}
\mathcal{A} ^\prime [\bw ,\balpha,\bbeta] \delta_q(j) = & \; \lim_{h\rightarrow0} h^{-1} 
\left( \mathcal{A}[ \bw + h\bdelta , \balpha,\bbeta] - \mathcal{A}[\bw,\balpha,\bbeta]\right) ,
\end{align}  
which implies 
\begin{align}
\mathcal{A} ^\prime [ \bw,\balpha,\bbeta] \delta_q = & \; \delta_q(j) \, \left( 
1+\log w_q(j))  +  u_q(j) + \beta_q +\alpha_j \genfrac{}{}{0pt}{1}{}{} \right)  . 
\label{EL-deriv1}
\end{align}
We are interested in `stationary' or `critical' points of the functional 
$\mathcal{A}$, which correspond to $\mathcal{A}^\prime[\bw,\balpha]\delta_q(j) = 0$ 
for all possible $\delta_q(j)$.  Note that here $\delta_q(j)$ are not completely arbitrary, 
there are constraints which $\delta_q(j)$ have to satisfy. 

If we perturb the terms involving $\alpha_j$ to $\alpha_j+h\hat\alpha_j$ and take 
the difference between $h\neq0$ and $h=0$, then we recover the conserved 
quantities in (\ref{Con1}) and (\ref{Con2}). In general, we have 
\begin{align}
\lim_{h\rightarrow0} h^{-1} \left( \mathcal{A}[\bw, \balpha + h \hat\balpha,\bbeta] 
- \mathcal{A}[\bw, \balpha,\bbeta] \right) = \hat\alpha_j C_j(\bw) , 
\end{align} 
and so by considering each component of  $\balpha$ variable in turn, 
we obtain each contraint $C_j=0$ (\ref{Con1}). 
The constraints (\ref{Con2}) are recovered by considering perturbations of 
$\bbeta=(\beta_+,\beta_0,\beta_-)$ to $\bbeta=(\beta_+ + \hat\beta_+, 
\beta_0 + \hat\beta_0, \beta_- + \hat\beta_-)$, that is, 
\begin{align}
\lim_{h\rightarrow0} h^{-1} \left( \mathcal{A}[\bw, \balpha,\bbeta + h \hat\bbeta] 
- \mathcal{A}[\bw, \balpha, \bbeta] \right) = \hat\beta_q C_q(w_q(j)) . 
\end{align} 
Setting this quantity to zero for arbitary $\hat\beta_q$ means the 
constraints $C_\pm=0=C_0$ are satisfied. 

The constraints (\ref{Con1}) and (\ref{Con2}) require that $\delta_q(j)$ satisfy 
\beq 
\delta_+(j) + \delta_0(j) + \delta_-(j) = 0 \, \;\; \forall j, \qquad  
\sum_{j=1}^N \delta_+(j) = 0 , \quad 
\sum_{j=1}^N \delta_0(j) = 0 , \quad 
\sum_{j=1}^N \delta_-(j) = 0 . 
\label{de-con}  \eeq 
To satisfy the last set of constraints, we replace $\delta_0(j)$ with 
$-\delta_+(j) - \delta_-(j)$,  this also satisfies the second constraint, 
provided that the first and third constraints hold.   Collecting terms 
in $\delta_+(j)$ and $\delta_-(j)$, equation (\ref{EL-deriv1}) implies 
\begin{align}
\mathcal{A} ^\prime [\bw,\balpha,\bbeta] \delta_+(j) = & \; \delta_+(j) \,\left[ \, 
\beta_+ - \beta_0 + u_+(j) - u_0(j) - \log w_+(j) + \log w_0(j) 
\genfrac{}{}{0pt}{1}{}{} \,\right] , \nn \\ 
\mathcal{A} ^\prime [\bw,\balpha,\bbeta] \delta_-(j) = & \; \delta_-(j) \,\left[ \, 
\beta_- - \beta_0 + u_-(j) - u_0(j) - \log w_-(j) + \log w_0(j) 
\genfrac{}{}{0pt}{1}{}{} \,\right]  .   \label{EL-d7}
\end{align}
Since we require $\mathcal{A} ` [w_q,\alpha_*] \delta_q = 0$ 
for all $\delta_\pm(j)$, we obtain the pair of equations 
\begin{align}
u_+(j) - u_0(j) = & \;  \log w_+(j) - \log w_0(j) - \beta_+ + \beta_0 , \\  
u_-(j) - u_0(j) = & \;  \log w_-(j) - \log w_0(j) - \beta_- + \beta_0 , 
\end{align}
which are quoted in the main text (\ref{eqmot4}). 

%----------------------------------------------------------------------------
\section{{Master equation approach} \label{res-sec}}
\setcounter{equation}{0}

The master equation approach refers to a methodology for describing 
the evolution of a stochastic system using variables to express the 
probability that a system is in a particular state at time $t$ \cite{krn}. 

We introduce a function which describes the probability of the system 
being in a certain state. Specifically, we let $G_\pm(j,k)$ be the probability 
that in positions $1,2,\ldots,j$ of the ordered list, there have been 
$k$ occurrences of the genetic state $\pm$.  We can write this 
formally as  $G_\pm(j,k) = \mathbb{P}[ W_\pm(j) = k]$. 

By conditioning the probability $\mathbb{P}[ W_+(j+1) = k+1]$
on the two possible states at $j$ (namely $k$ or $k+1$), that is 
\beq
\mathbb{P}[ W_+(j+1) = k+1] = \mathbb{P}[ W_+(j) = k] w_+(j+1) 
+ \mathbb{P}[ W_+(j) = k+1] (1-w_+(j+1)) , 
\eeq
we obtain a recurrence relation for $G_\pm$. 
\begin{align}
G_+( j+1,k+1) = & \; G_+(j,k) w_+(j+1) + (1-w_+(j+1)) G_+(j,k+1) , \nn \\ 
G_-( j+1,k+1) = & \; G_-(j,k) w_-(j+1) + (1-w_-(j+1)) G_-(j,k+1) , 
\label{Gmaster}
\end{align}
We note that the equations (\ref{Gmaster}) are almost identical, only 
differing in the $\pm$ subscripts, so any analysis of the equations can 
be undertaken on a general case, and the results obtained will be 
applicable in both the $\pm$ cases, hence, below, we ignore the subscripts. 
Clearly $k\geq0$ and $k \leq j$; thus this system has to be solved subject 
to the boundary conditions $G_\pm(j,-1)=0$ and $G_\pm(j,j+1)=0$, and 
the `initial' condition $G(0,0)=1$; here we treat $j$ as a time-variable, 
with the region $0<j<N$ being the range of interest. 

In the case of large $N$, a continuum limit argument can be used to 
determine the spread of the probability distributions $G(j,k)$, and show 
that this has the form of a Gaussian distribution.  We define a small 
parameter $h \ll1$, by $h = 1/N$, and a continuum limit of the probability by 
\beq
G(j,k) = h \, \tilde G(\tau,y), \quad \mbox{ where } \;\;\; 
\tau = h(j-\half), \quad y=h(k+\half) ,  \label{GGtilde}
\eeq
The scaling $G=h\tilde G$ is introduced so that the condition $\sum_k G(j,k) 
=1$ for all $j$ is transformed into $\int\tilde G \, \dd y = 1$ for all $\tau$. 

Following (\ref{Gmaster}), the governing equation for $\tilde G(\tau,y)$  is 
\beq
\tilde G(\tau+\half h, y + \half h ) = \tilde w (\tau+\half h) \tilde G(\tau-\half h , y-\half h) 
+ (1 - \tilde w(\tau+\half h)) \tilde G (\tau-\half h , y + \half h ) , 
\label{Gtilde-diffeq}
\eeq
where $\tilde w(\tau) = w(j+\half)$.  Shifts of $j,n$ in the definition of 
continuum limit amount to a choice about which point to perform a Taylor 
series expanion, and simplify later analysis. The continuum version of 
the cumulative distribution is defined by $\tilde{W}(y) = \frac{1}{N} W(j)$, 
so that $\tilde W(0)=0$ and $\tilde W(1)=w^{(0)}$. 
Since $w(j) = W(j) - W(j-1)$ we also have $\tilde w(\tau) = \tilde W(\tau) 
- \tilde W(\tau-h) \approx h \tilde W'(\tau)$ and $\tilde W = h W$ 
together with $\tilde W'(\tau) = \tilde w(\tau)$. 
Taking Taylor series of (\ref{Gtilde-diffeq}) in $G$, upto and including 
terms of $\mathcal{O}(h^2)$, we find the PDE 
\beq
\pad{\tilde{G}}{\tau} + \tilde w(\tau) \pad{\tilde{G}}{y} + 
\half h (1-\tilde{w}(\tau)) \paddd{\tilde{G}}{y}{\tau} = 0. 
\label{G-PDE}
\eeq
We are interested in the solution on the domain $0<\tau<1$ and $0<y<\tau<1$ 
with $\tilde{G}(\tau,0)=0=\tilde{G}(\tau,\tau)=0$, and $G(0,y)=\delta(y)$ 
(this being the Dirac delta function).  

The leading order terms of (\ref{G-PDE}) are $\tilde G_\tau = -\tilde w(\tau) 
\tilde G_y$, which gives the leading order travelling wave solution $\tilde G = 
\tilde{G}(y-\tilde W(\tau),\tau)$.   At any particular value of $\tau$, the mean of 
the distribution $G$ is given by $W(j) = N \tilde W(y)$.  We define a new 
variable $z$ for this quantity, and seek a solution of the form 
\beq
\tilde G(\tau,y) = \frac{1}{\sqrt{h}} \check G (\tau,z) , \quad 
z = \frac{ y - \tilde W(\tau) }{\sqrt{h}} . \label{GtGh} 
\eeq 
Again the scaling between $\tilde G$ and $\check G$ ensures $\int \tilde G \, 
\dd y = 1$ is mapped to $\int \hat G \, \dd z=1$ for all $\tau$. Returning to the 
second-order expansions of (\ref{G-PDE}), we obtain an equation of 
Fokker-Planck type 
\beq
\pad{ \hat G}{\tau} = \half \tilde w(\tau) (1-\tilde w(\tau) ) \padd{ \hat G}{z} , 
\eeq
which has the solution 
\beq
\hat G(\tau,z) = \frac{ \ee^{-z^2/ 4 s(\tau) } }{ 2 \sqrt{\pi s(\tau)}} , \quad \mbox{
where} \;\;\; \frac{\dd s(\tau)}{\dd \tau} = \half \tilde w(\tau) (1-\tilde w(\tau)) . 
\eeq
Inverting the transformations using  (\ref{GtGh}) and (\ref{GGtilde}), we obtain 
\beq
G_+(j,k) = \frac{1}{2\sqrt{\pi N s_+(j)}} \exp\left( - \frac{ (k-W_+(j))^2 }{ 4 N s_+( j )} \right) , 
\qquad s_+(j) = \frac{1}{2N} \sum_{i=1}^j w_+(i) [ 1-w_+(i) ] . 
\eeq
This shows how far from the expected value we would expect to see 
stochastic fluctuations; similar formula hold for $G_-,G_0$, with 
corresponding formulae for $s_-,s_0$ in terms of $w_-(i),w_0(i)$.  
Since $\theta_\pm(j) = W_\pm(j) - j w_\pm^{(0)} = W_\pm(j) - j N_\pm / N$, 
the variance of $\theta_\pm(j)$ will be the same as the variance in $W_\pm(j)$. 

%---------------------------------------------
% \subsection{{\bf Null hypothesis \& $p$-values}}

The null hypothesis corresponds to the assumption that the gene has no effect 
on the phenotype, which is equivaelent to the case of random allocation 
discussed at the end of  Section \ref{math-sec}. Here we assume that 
$\tilde{w}=w^{(0)}$, hence $W(j)=j w^{(0)}$, $s (j)=w^{(0)}(1-w^{(0)})j/2N$ and so 
\beq
G_+(j,k) 
= \frac{1}{\sqrt{2\pi j w_+^{(0)} (1-w_+^{(0)})}} \, \exp 
\left( \frac{ -(k-jw_+^{(0)})^2 }{ 2 j w_+^{(0)} (1-w_+^{(0)} )} \right) 
% = \frac{1}{\sqrt{2\pi j w_+^{(0)} (1-w_+^{(0)})}} \, \exp 
% \left( \frac{ -\theta_+^2 }{ 2 j w_+^{(0)} (1-w_+^{(0)} )} \right) 
. \label{appB-Gerf} \eeq
The maximal variance would occur at $j=N/2$ since we require 
$G(0,k)=\delta_{k,0}$ and $G(N,k) = \delta_{k,N_q}$. 
The combination $k-j w^{(0)}$ corresponds to our $\theta$ variable. 

Whilst this approach works well for the early stages in the list, 
$1 \leq j \leq N/2$, for later locations (as $j \nearrow N$), the 
distibution should reduce in variance, and converge to the single 
point $G_q(N,k) = \delta_{k,N_q}$.  This PDE approach is not 
able to describe this type of behaviour. % \textinterrobang \ 
One could work back from this `initial' condition and aim to match the 
variances of the two continuum solutions: one from $\tau$ increasing 
from zero and the other with $\tau$ decreasing from 1. 
The difference between (\ref{appB-Gerf}) and the distributions 
calculated in Section \ref{pval-sec} is accounted for by this effect. 

%--------------------------------
\section{{\bf Phenotypic-dependent distributions}  \label{pheno-dep-sec} }
\setcounter{equation}{0}

In Section \ref{deriv-sec}, we considered just the {\em location} of an individual 
in an ordered list ($1\leq j \leq N$);  in some contexts it may make more 
sense to think of the distributions $u_q(j)$, $W_q(j)$, and $\theta_q(j)$-paths as 
functions of {\em phenotype value}, $\Omega$, (where $\Omega$ is, 
for example, height).  To model this alternative way of thinking, we 
define $\tilde W_q(\Omega)$ by 
\begin{align} 
\label{C1}
\tilde W_q(\Omega) = &\; \mbox{number of individuals of genetic state $q$ 
with phenotype $<\Omega$}, \\ 
\tilde W_q^{(0)}(\Omega) = & \; \mbox{expected number of individuals 
of genetic state $q$ with phenotype $<\Omega$} \nn \\  \nn 
& \; \; \mbox{(the `random' configuration, i.e.~the average over 
all possible arrangements). }
\end{align} 
and generalise $\theta_q(j)$ to 
\beq
\tilde\theta_q(\Omega) = \tilde W_q(\Omega) - \tilde W_q^{(0)}(\Omega) , 
\qquad q \in \{+1,0,-1 \} .  
\label{C2}
\eeq
In general, since the cumulative distribution of $\Omega$, $P(\Omega)$, 
is unknown, there is no simple expression for $\tilde W_q^{(0)}(\Omega)$ 
similar to (\ref{WW0-def}).  We define $p(\Omega) = P'(\Omega)$ as the 
probability density function of phenotype. 

To relate this phenotypic-dependent formulation with the original (list-based), 
we write $\Omega(j)$ as the phenotype of individual $j$, and make use of 
the relations
\beq
W_q(j) = \tilde W_q(\Omega(j)) , \quad 
W_q^{(0)}(j) = \tilde W_q^{(0)}(\Omega(j)) , \quad 
\theta_q(j) = \tilde \theta_q(\Omega(j)) .  
\label{WW-WWtilde}
\eeq
\qb{Similarly, the phenotype-dependent fields $\tilde u_q(\Omega)$ can 
be obtained from $\tilde u_q(\Omega(j)) = u_q(j)$; and this field 
can be related to the local genotype probability density, $\tilde w_q(\Omega)$ 
using an extension to the formula} (\ref{eqmot4}), \qb{namely} 
\begin{align}
\tilde u_+(\Omega) - \tilde u_0 (\Omega) = & \; \log \tilde w_+(\Omega) 
	- \log \tilde w_0(\Omega) - \beta_+ + \beta_0 , \nn \\ 
\tilde u_-(\Omega) - \tilde u_0 (\Omega) = & \; \log \tilde w_-(\Omega) 
	- \log \tilde w_0(\Omega) - \beta_+ + \beta_0 . 
\end{align}
In general, the distributions of the phenotypes for the three genotypes 
could be different, that is, we have distinct probability density functions 
$p_+(\Omega),p_0(\Omega),p_-(\Omega)$ with corresponding 
cumulative distributions $P_+(\Omega),P_0(\Omega),P_-(\Omega)$. 

If we make the assumption that the three distributions are almost equal, 
that is 
\beq
P_q(\Omega) \approx P(\Omega) + \mathcal{O}(h) 
\quad \mbox{with} \quad h \ll1 , 
\label{C4}
\eeq 
then we can construct the quantity $\tilde w_q(\Omega)$ 
akin to $w_q(j)$ (\ref{wq-def}). 
The expected values of the order statistics $\Omega(j)$ are given by 
$\mathbb{E}[ P(\Omega(j)) ]= (2j-1)/2N$;  taking the difference of this 
with respect to $j$ gives  
\beq
\mathbb{E} \left[ P(\Omega(j)) - P(\Omega(j-1) \right] \approx 
\mathbb{E} \left[ P'(\Omega) \left(\Omega(j) - \Omega(j-1) \right) \right] 
= p(\Omega) \Delta \Omega = \frac{1}{N} , 
\label{DOmega}
\eeq
where $\Omega \in (\Omega(j-1),\Omega(j))$. Since the derivative 
of the cumulative distribution function $P(\Omega)$ is the density 
function $p(\Omega)$, and $N\gg1$ we have 
\beq
\Delta \Omega = \Omega(j) - \Omega(j-1) \approx \frac{1}{ N p(\Omega) } . 
\label{DOm2}
\eeq
This describes the expected separation between individual's phenotypes 
in the sample $\Gamma$ (\ref{Gamma-def}). 

Combining (\ref{wq-def}), (\ref{WW-WWtilde}), (\ref{DOmega}) noting that, at 
leading order, the distribution of phenotype states is given by $P(\Omega)$, 
we obtain 
\begin{align}
w_q(j) = & \; W_q(j) - W_q(j\!-\!1) 
= \tilde W_q(\Omega(j)) - \tilde W_q(\Omega(j\!-\!1)) \nn \\ 
\approx & \; \frac{\dd \tilde W_q}{\dd \Omega} ( \Omega(j) - \Omega(j\!-\!1) ) 
= \frac{1}{N p(\Omega)} \frac{\dd \tilde W_q}{\dd \Omega} 
= (\Delta \Omega) \frac{\dd \tilde W_q}{\dd \Omega} . 
\label{C7}
\end{align} 
In the case of the expectation of the random configuration,  
this calculation amounts to a consistency condition 
\begin{align}
w_q^{(0)} = & \; W_q^{(0)}(j) - W_q^{(0)}(j-1) 
= \tilde W_q^{(0)}(\Omega(j)) - \tilde W_q^{(0)}(\Omega(j-1)) \nn \\ 
\approx & \; \frac{\dd \tilde W_q^{(0)} }{\dd \Omega} ( \Omega(j) - \Omega(j\!-\!1) ) 
\approx \frac{N_q}{N p(\Omega)} \frac{\dd P}{\dd \Omega} = \frac{N_q}{N} . 
\label{C8}
\end{align}
Thus, it is natural to define 
\beq
\tilde w_q(\Omega) = \frac{\dd \tilde W_q}{\dd \Omega} \Delta\Omega 
= \frac{\dd \tilde W_q}{\dd \Omega} \frac{1}{N p(\Omega) },  
\label{C9}
\eeq
so that we have $\tilde w_q = w_q^{0}$ in the random case. 
The interpretation of the paths $(\theta_+,\theta_-)$ and 
$(\tilde\theta_+,\tilde\theta_-)$ follow the development of a 
mathematical model which relates phenotype to genotype via a `field', 
which is determined using the distributions $W_q,w_q,\tilde W_q,\tilde w_q$.

% Discuss $\tilde u_q(\Omega)$  % 16.9 

%----------------------------------------------------------------------------

%\renewcommand{\baselinestretch}{0} 
\footnotesize

\setcounter{tocdepth}{2}
%\tableofcontents

\end{document}